\documentclass{aa}  

\usepackage{graphicx}
\usepackage{txfonts}
\usepackage[flushleft]{threeparttable}
\usepackage{tabularx}
\def\arraybackslash{\let\\\tabularnewline} 

\usepackage{xcolor}

\usepackage{gensymb}

\usepackage{soul}

\usepackage[labelformat=simple,justification=centering]{subcaption}
\renewcommand\thesubfigure{(\alph{subfigure})}

\begin{document}

   \title{Gas temperature structure across transition disk cavities}

   \author{M. Leemker
          \inst{1}
          \and
          A. S. Booth\inst{1}
          \and
          E. F. van Dishoeck\inst{1, 2}
          \and
          A. F. P\'{e}rez-S\'{a}nchez\inst{1} 
          \and
          J. Szulágyi\inst{3}
          \and
          A. D. Bosman\inst{4}
          \and
          S. Bruderer\inst{2}
          \and
          S. Facchini\inst{5,6}          
          \and
          M. R. Hogerheijde\inst{1,7} 
          \and
          T. Paneque-Carre{\~n}o\inst{1,5}
          \and
          J. A. Sturm\inst{1}}
          
   \institute{Leiden Observatory, Leiden University, P.O. box 9513, 2300 RA Leiden, The Netherlands\\
              \email{leemker@strw.leidenuniv.nl}
         \and
         	 Max-Planck-Institut f\"ur Extraterrestrische Physik, Giessenbachstrasse 1, 85748 Garching, Germany 
         \and
             Institute for Particle Physics and Astrophysics, ETH Z\"{u}rich, 8093 Z\"{u}rich, Switzerland
    	 \and 
    	    Department of Astronomy, University of Michigan, 323 West Hall, 1085 S. University Ave., Ann Arbor, MI 48109, USA         
         \and
             European Southern Observatory, Karl-Schwarzschild-Str. 2, 85748 Garching, Germany 
        \and
        	Universit\`a degli Studi di Milano, via Giovanni Celoria 16, 20133 Milano, Italy
        \and 
    	    Anton Pannekoek Institute for Astronomy, University of Amsterdam, Science Park 904, 1090 GE Amsterdam, The Netherlands
             }

   \date{Received XXX; accepted YYY}

  \abstract
{Most disks observed at high angular resolution show signs of substructures like rings, gaps, arcs, and cavities in both the gas and the dust. To understand the physical mechanisms responsible for these structures, knowledge about the gas surface density is essential. This, in turn, requires information on the gas temperature.}
{The aim of this work is to constrain the gas temperature as well as the gas surface densities inside and outside the mm-dust cavities of two transition disks: LkCa15 and HD~169142, with dust cavities of 68~AU and 25~AU, respectively. }
{We use some of the few existing ALMA observations of the $J=6-5$ transition of $^{13}$CO together with archival $J=2-1$ data of $^{12}$CO, $^{13}$CO and C$^{18}$O. The ratio of the $^{13}$CO $J=6-5$ to the $J=2-1$ transition is used to constrain the temperature and is compared with that found from peak brightness temperature of optically thick lines. The spectra are used to resolve the innermost disk regions to a spatial resolution better than the beam of the observations. Furthermore, we use the thermochemical code DALI to model the temperature and density structure of a typical transition disk as well as the emitting regions of the CO isotopologues.}
{The $^{13}$CO $J=6-5$ and $J=2-1$ transitions peak inside the dust
cavity in both disks, indicating that gas is present in the dust
cavities. The kinematically derived radial profiles show that the gas
is detected down to 10 and $5-10$~AU, much further in than the dust
cavities in the LkCa15 and HD~169142 disks, respectively. For LkCa15,
the steep increase towards the star in the $^{13}$CO $J=6-5$ transition,
in contrast to the $J=2-1$ line, shows that the gas is too warm to
be traced by the $J=2-1$ line and that molecular excitation is
important for analysing the line emission. Quantitatively, the $6-5/2-1$
line ratio constrains the gas temperature in the emitting layers
inside the dust cavity to be up to 65~K, warmer than in the outer disk at
20$-$30~K.  For the HD~169142, the lines are optically thick,
complicating a line ratio analysis. In this case, the peak brightness
temperature constrains the gas in the dust cavity of HD~169142 to
be 170~K, whereas that in the outer disk is only 100~K. The data
indicate a vertical structure in which the $^{13}$CO $6-5$ line emits from a
higher layer than the $2-1$ line in both disks, consistent with
exploratory thermo-chemical DALI models. Such models also show that a
more luminous central star, a lower abundance of PAHs and the absence
of a dusty inner disk increase the temperature of the emitting layers
and hence the line ratio in the gas cavity. The gas column density in the LkCa15 dust cavity drops by a factor $>2$ compared to the outer disk, with an additional drop of an order of magnitude inside the gas cavity at 10~AU. In the case of HD~169142, the gas column density drops by a factor of 200$-$500 inside the gas cavity. }
{
The gas temperatures inside the dust cavities steeply increase towards the star and reach temperatures up to 65~K (LkCa15) and 170~K (HD~169142) on scales of $\sim$15-30~AU, whereas the temperature gradients of the emitting layers in the outer disks are shallow with typical temperatures of 20 $-$
30 and 100~K, respectively. The deep drop in gas column density
inside the HD~169142 gas cavity at <10~AU could be due to a massive companion
of several M$_{\mathrm{J}}$, whereas the broad dust-depleted gas region from 10-68~AU for LkCa15 may imply several lower mass planets. This work demonstrates that knowledge of the gas temperature is important to determine the gas surface density and thus whether planets, and if so what kind of planets, are the most likely carving the dust cavities.}


   \maketitle

\section{Introduction}
High angular resolution observations of protoplanetary disks show structures in both gas and dust (e.g., \citealt{Andrews2010, vanderMarel2015, Fedele2017, Andrews2018, Huang2018, Long2018, Oberg2021}; see \citealt{Andrews2020} for review). One favoured mechanism that can explain most of these substructures is planet-disk interactions ($>$ 20 M$_{\mathrm{Earth}}$) (e.g. \citealt{Bryden1999, Zhu2014, Dipierro2015, Rosotti2016, Dong2018, Zhang2018DSHARP, Szulagyi2018, Binkert2021}). If these structures are caused by embedded planets, these gaps should be deep in both gas and dust (at least a factor of 10 depletion in the gas). However, the structures could also caused by a change in the temperature that is unrelated to a potential embedded planet. Therefore, knowledge of the temperature profile across these structures is needed to distinguish truly empty gaps from emission drops related to temperature decreases.

There are only a few cases where young exoplanets have been caught in the act of formation. Two young planets and their disks have recently been directly imaged in the central dust cavity of the PDS~70 disk \citep{Keppler2018, Haffert2019, Isella2019, Benisty2021} and three more have been located in gaps in the HD~163296 and HD~97048 disks based on localized azimuthal perturbations in the disk velocity structure traced by CO kinematics (\citealt{Teague2018, Pinte2019, Teague2019, Izquierdo2021}). Two of these disks are typical transition disks with large mm-dust cavities, supporting the expectation that planets are present in most (transition) disks. 
The presence of massive planets ($>$ several M$_{\mathrm{Jup}}$) in a number of other disks including those around HD~100546 and HD~169142 has been suggested, but confirmation by direct imaging is still lacking (e.g., \citealt{Quanz2013HD100546, Quanz2013HD169142, Osorio2014, Reggiani2014, Walsh2014, Currie2017, Follette2017, Rameau2017, Casassus2019, Toci2020}). Similarly, planets have been proposed and refuted in the LkCa15 disk \citep{Kraus2012, Sallum2015, Currie2019}.
Other explanations for the structures seen in disks include disk winds, internal photoevaporation, opacity variations, chemical effects, and magneto-hydrodynamics (e.g., \citealt{Clarke2001, Birnstiel2015, Zhang2015, Flock2015, Simon2016}). 
Spatially resolved observations of gas in and near a gap or cavity can distinguish between these scenarios. In particular, the presence of massive planets should result in deep gas cavities (gas depleted by at least a factor of 10) that are somewhat smaller than the deep dust cavities.

In this work, we focus on transition disks with large cavities as they are more easily resolved by current ALMA observations. Fully resolving the cavity with multiple beams is crucial to derive the properties of the gas and dust in the cavities \citep{Bruderer2013, Szulagyi2018}. The gas shows a different morphology than the dust, with the gas cavity indeed typically smaller than that in the dust (e.g., \citealt{Bruderer2014, Perez2015, vanderMarel2015, vanderMarel2016, Woelfer2021}). However, observations do not trace the column density of the gas and dust directly. A gas cavity seen in molecular line emission can be due to an actual decrease in the surface density, but it can also be caused by a drop in the gas temperature in that region of the disk or a combination of both \citep{Facchini2017, Facchini2018}. Therefore, the temperature needs to be known to accurately derive the surface density in the gas cavity.

The gas temperature inside and outside the cavity is controlled by a number of heating and cooling processes including the photoelectric effect on PAHs and small grains exposed to UV radiation, gas-grain coupling  and atomic and molecular line emission (e.g. \citealt{Bakes1994, Kamp2001, Weingartner2001, Gorti2004, Woitke2009, Bruderer2012, Bruderer2013}).
A crucial ingredient is the local UV field. The radiation field at each point in the disk depends on a number of parameters: the stellar type of the host star, the UV luminosity from accretion of material onto the star, the abundance of dust and PAHs and their properties, and the presence of an inner disk that shields radiation from the host star. 
PAHs that heat the gas through the photoelectric effect and dusty inner disks are commonly observed in transition disks \citep{Habart2006, Brown2007, Geers2007a, Geers2007b, Merin2010, Isella2019, Perez2019, Facchini2020, Francis2020}. 
Detailed modelling by \citet{Jonkheid2006, Woitke2009, Bruderer2012, Bruderer2013, Facchini2017, Facchini2018, Alarcon2020, Rab2020} shows that the temperature of the gas changes when a gap or cavity is introduced in the model. Whether the gas becomes warmer or colder than the gas in the full disk model depends on parameters like the grain sizes and on the gas-to-dust ratio. A planet in the gap can locally increase the temperature by several tens of K \citep{Cleeves2015, Szulagyi2018}.

Pre-ALMA the gas temperature in the emitting layers has been derived by comparing observations of various CO lines, up to high $J_{\mathrm{u}}\geq 14$-lines observed with ground based telescopes and \textit{Herschel} (e.g. \citealt{vanZadelhoff2001, Sturm2010, Bruderer2012, Dent2013, Fedele2013, Meeus2012, Green2013, Meeus2013}). These observations mainly  probe the temperature in intermediate disk layers ($z/r\sim 0.5$) at 10-200~AU from the central star \citep{Fedele2016}. 

With the advent of ALMA, spatially resolved observations have been used together with detailed modelling of individual disks with large dust cavities (e.g., \citealt{Bruderer2012, vanderMarel2015, vanderMarel2016, Kama2016, Schwarz2021}) and without large dust cavities (e.g., \citealt{Schwarz2016, Pinte2018temperature, Calahan2021}). 
The temperature of the (outer) disk midplane is derived from the location of the snowlines of major species such as H$_2$O and CO \citep{Mathews2013, Qi2013, vantHoff2017, vantHoff2018methanol, Qi2019, Leemker2021}. Additionally rotational diagrams are used to derive the temperature of intermediate layers in disks (e.g., \citealt{Schwarz2016, Loomis2018, Pegues2020, Jeroen2021}).
In this work, we study the rarely observed $J=6-5$ transition of $^{13}$CO ($E_{\mathrm{u}} = 111.1$~K) in two transition disks: LkCa15 and HD~169142 with spatially resolved ALMA data. The combination of this high transition with the more commonly observed $^{13}$CO $J=2-1$ ($E_{\mathrm{u}} = 15.9$~K) transition, allows us to study the temperature and column density in the cavities of two transition disks and link it to the rest of the disk. The very high resolution of the $^{13}$CO $J=6-5$ data provides a unique opportunity to study the gas temperature in these disks in detail.

The sources and observations are discussed in Section~\ref{sec:obs}. Next, the continuum and molecular line emission across the cavities of LkCa15 and HD~169142 are presented in Section~\ref{sec:results}. The relevant equations for the temperature analysis are summarized and applied to the data in Section~\ref{sec:ana}, where we derive the temperature, column density and optical depth using CO isotopologues. We put the derived temperature structure in context using a representative thermochemical model in Section~\ref{sec:dalisection}. Finally, we discuss and summarize our findings in Section~\ref{sec:disc2} and \ref{sec:concl}, respectively.

\section{Observations} \label{sec:obs}

\begin{table*}[]
\begin{threeparttable}
    \centering
	\caption{Source properties }  
	\begin{tabularx}{\linewidth}{p{0.15\columnwidth}p{0.19\columnwidth}p{0.21\columnwidth}p{0.06\columnwidth}p{0.11\columnwidth}p{0.08\columnwidth}p{0.1\columnwidth}p{0.1\columnwidth}p{0.1\columnwidth}p{0.1\columnwidth}p{0.07\columnwidth}p{0.07\columnwidth}p{0.12\columnwidth}}    
    \hline\hline
      \\[-0.7em] 
      \centering\arraybackslash Disk & \centering\arraybackslash RA $J2016$ & \centering\arraybackslash Decl. $J2016$ & \centering\arraybackslash $M_*$ (M$_{\odot}$) & \centering\arraybackslash $\dot{M}$ (M$_{\odot}$ yr$^{-1}$) & \centering\arraybackslash $L_*$ (L$_{\odot}$) & \centering\arraybackslash spectral type & \centering\arraybackslash age (Myr) & \centering\arraybackslash distance (pc) &\centering\arraybackslash  $v_{\mathrm{sys}}$ (km s$^{-1}$) & \centering\arraybackslash incl. ($\degree$) & \centering\arraybackslash PA ($\degree$) & \centering\arraybackslash Refs.  \\ \hline
        \\[-0.7em]
LkCa15 & 	04:39:17.79 & 22:21:03.39  & 1.3 & $10^{-9.2}$~& 1.1  & K3 & $\sim$5 & 157.2 & 6.25 & 50  & 62 & 1-5 \\   
HD~169142 &  18:24:29.78 & -29:46:49.33 & 1.7 & $10^{-7.4}$~ & 10 & F1 & $\sim$6 & 114.9 & 6.9 & 13 & 5 & 1,6-13\\       \hline        
    \end{tabularx}
    \begin{tablenotes}
      \small
      \item \textbf{References.} ${(1)}$~\citealt{GaiaDR22018}, ${(2)}$~\citealt{Donati2019}, ${(3)}$~\citealt{Wolk1996}, ${(4)}$~this work, ${(5)}$~\citealt{Facchini2020}, ${(6)}$~\citealt{Blondel2006}, ${(7)}$~Following \citealt{Fedele2017}, based on the optical and UV-extinction \citep{Malfait1998} and the new Gaia distance, ${(8)}$~\citealt{Gray2017}, ${(9)}$~\citealt{Grady2007}, ${(10)}$~\citealt{Fedele2017}, ${(11)}$~\citealt{Raman2006}, ${(12)}$~\citealt{Panic2008}, and ${(13)}$~\citealt{Guzman-Diaz2021}. 
    \end{tablenotes}
    \label{tab:sources}
\end{threeparttable}
\end{table*}

\begin{table*}[]
\begin{threeparttable}
    \centering
	\caption{Molecular line and continuum observations used in this work}    
     \begin{tabularx}{\linewidth}{p{0.14\columnwidth}p{0.25\columnwidth}p{0.13\columnwidth}p{0.1\columnwidth}p{0.08\columnwidth}p{0.37\columnwidth}p{0.07\columnwidth}p{0.07\columnwidth}p{0.22\columnwidth}p{0.18\columnwidth}}    
    \hline\hline \noalign {\smallskip}
       \centering\arraybackslash Disk & \centering\arraybackslash Transition &  \centering\arraybackslash Int. flux$^{(1)}$ (Jy km~s$^{-1}$) & \centering\arraybackslash Channel width (km~s$^{-1}$) & \centering\arraybackslash Channel rms (mJy beam$^{-1}$) & \centering\arraybackslash Beam & \centering\arraybackslash MRS$^{(2)}$ & \centering\arraybackslash Robust  & \centering\arraybackslash Project code & \centering\arraybackslash PI  \\ \hline
        \\[-0.7em]
       LkCa15    & $^{12}$CO $J=2-1$ & $16\pm 2$ &  0.04 & 8.9 & $0\farcs34 \times 0\farcs25\ (-9.7\degree)$ & $3\farcs7$ & $-2.0$ &  2018.1.01255.S & M.\,Benisty \\        
                 & $^{13}$CO $J=2-1$ & $1.4\pm0.2$ & 0.17 & 4.4 & $0\farcs17 \times 0\farcs13 (-13.7\degree)$ & $0\farcs5$ & nat.$^{(5)}$ & 2018.1.00945.S & C. Qi \\ 
                 & $^{13}$CO $J=2-1$ & $5.8\pm 0.6$ & 0.33 & 3.1 & $0\farcs35 \times 0\farcs25\ (26.6\degree)$ & $4\farcs0$ & $-2.0$ & 2018.1.00945.S & C. Qi \\ 
                 & $^{13}$CO $J=6-5$ & $7.5\pm 0.8$ & 0.5 & 2.8 & $0\farcs08 \times 0\farcs05\ (-18.6\degree)$ & $1\farcs2$ & $0.5^{(6)}$ &  2017.1.00727.S & J. Szulágyi \\
                 & $^{13}$CO $J=6-5$ & $8.2\pm 0.9$ & 0.44 & 25 & $0\farcs32\times 0\farcs30\ (-31.2\degree)$ & $2\farcs1$ & $0.5$ &  2017.1.00727.S & J. Szulágyi \\
                 & C$^{18}$O $J=2-1$ & $1.0\pm 0.1$ & 0.33 & 3.9 & $0\farcs36 \times 0\farcs27\ (26.6\degree)$ & $4\farcs0$ & $-2.0$ &  2018.1.00945.S & C. Qi \\ 
                 & 0.45 mm cont. & &  & 0.16 & $0\farcs071\times 0\farcs048\ (-23.5\degree)$ & $1\farcs2$ & 0.5 &  2017.1.00727.S & J. Szulágyi  \\                  
        \\[-0.5em]         
        HD~169142 & $^{12}$CO $J=2-1$ & $9.3\pm 0.9$ & 0.17 & 1.7 & $0\farcs05\times 0\farcs029\ (78.8\degree)$ & $0\farcs6$ & 0.5 & 2016.1.00344.S & S. P\'{e}rez$^{(3)}$ \\
                 & $^{13}$CO $J=2-1$ & $4.3\pm 0.4$ & 0.17 & 2.3 & $0\farcs054\times 0\farcs035\ (78.4\degree)$ & $0\farcs6$ & 0.5 & 2016.1.00344.S & S. P\'{e}rez$^{(3)}$\\
                 & $^{13}$CO $J=6-5$ & $6.1\pm 0.7$ & 0.44 & 12 & $0\farcs055\times 0\farcs054\ (86.2\degree)$ & $0\farcs6$ & 0.0 & 2017.1.00727.S & J. Szulágyi \\
                 & C$^{18}$O $J=2-1$ & $2.0\pm 0.2$ & 0.17 & 1.4 & $0\farcs052\times 0\farcs031\ (78.6\degree)$ & $0\farcs6$ & 0.5  & 2016.1.00344.S & S. P\'{e}rez$^{(3)}$\\
                 & 0.45 mm cont. &  & & 0.52 & $0\farcs038\times 0\farcs037\ (-72.0\degree)$ &  $0\farcs6$ & 0.5$^{(7)}$ &  2017.1.00727.S & J. Szulágyi  \\                  
                 & 1.3 mm cont. &  &  & 0.022 & $0\farcs036\times 0\farcs018\ (73.0\degree)$ & $0\farcs6$ & $-0.5$ &  2016.1.00344.S & S. P\'{e}rez$^{(3)}$ \\ 
       \hline        
    \end{tabularx}
    \begin{tablenotes}
      \small
      \item \textbf{Notes.} $^{(1)}$ Disk integrated flux calculated from the region within the Keplerian mask. The error includes the 10\% absolute flux calibration error, but not any uncertainty due to resolved out large scale emission. $^{(2)}$ The maximum recoverable scale (MRS) is estimated as $\sim0.983\lambda/L_{\mathrm{5}}$, with $\lambda$ the wavelength and $L_{\mathrm{5}}$ the 5$^{th}$ percentile baseline of the configuration. $^{(3)}$ Data first presented by \citet{Perez2019}. $^{(4)}$ Data first presented in \citet{Facchini2020}. $^{(5)}$ Natural weighting with a $0\farcs15$ $uv$-taper. $^{(6)}$ ALMA image for which high and intermediate spatial resolution ALMA data are combined. $^{(7)}$ Product data. 
    \end{tablenotes}
    \label{tab:obs}
\end{threeparttable}
\end{table*}

In this paper we analyse ALMA observations of CO isotopologues in the LkCa15 and HD~169142 disks. An overview of the stellar parameters is given in Table~\ref{tab:sources} and an overview of the observations is presented in Table~\ref{tab:obs}. The sources and data reduction are discussed in this section. 

\subsection{The sources}

LkCa15 is a transition disk with a large dust cavity up to 68~AU radius. The system is located at $157.2\pm 0.7$~pc \citep{GaiaDR22018}. The disk surrounding the K3 type T~Tauri star LkCa15 is inclined by $\sim50-55\degree$ (\citealt{Wolk1996, vanderMarel2015, Facchini2020}). High resolution ALMA Band~6 observations reveal that, in addition to the cavity, the dust is highly structured with multiple narrow rings observed at high resolution \citep{Facchini2020}. Small grains, traced by scattered light observations, have been seen out to $\sim30$~AU radius inside the mm-dust cavity, suggesting that gas is present in this disk region \citep{Thalmann2016, Oh2016}. Furthermore, ALMA observations have revealed that the dust cavity is not completely devoid of mm grains as the continuum intensity in the dust cavity peaks at $73.7\pm 6.9~\mu$Jy~beam$^{-1}$ \citep{Facchini2020}. Here we present the first ALMA observations of a gas cavity in the LkCa15 disk. Previous observations by \citet{vanderMarel2015} do not have the spatial resolution to resolve this gap in the gas. Furthermore, CO ro-vibrational lines at $5~\mu$m also indicate a low CO gas column in the inner 0.3~AU of the LkCa15 disk \citep{Salyk2009}.

The dust in HD~169142 is highly structured similar to the LkCa15 disk \citep{Fedele2017, Perez2019}. Yet, HD~169142 is a Herbig F1 star \citep{Gray2017} with a mass of $1.7$~M$_{\odot}$ \citep{Blondel2006} at $114.9\pm 0.4$~pc \citep{GaiaDR22018} with its surrounding disk seen almost face-on ($i=13\pm 1\degree$, \citealt{Raman2006, Panic2008}). A small ($\sim0.3$~AU) and variable NIR inner dust disk has been observed in HD~169142 \citep{Wagner2015, Chen2019}. This inner dust disk has also been detected in high resolution ALMA observations \citep{Perez2019}.
The HD~169142 disk not only has structures in the dust, but also in the gas with rings and gaps. A gas cavity is seen in CO isotopologues, and DCO$^+$ is found to have an inner component and a ring at $\sim50-230$~AU \citep{Fedele2017, Carney2018}.

\subsection{Data}

\begin{figure*}
        \centering
        \includegraphics[width=0.9\textwidth]{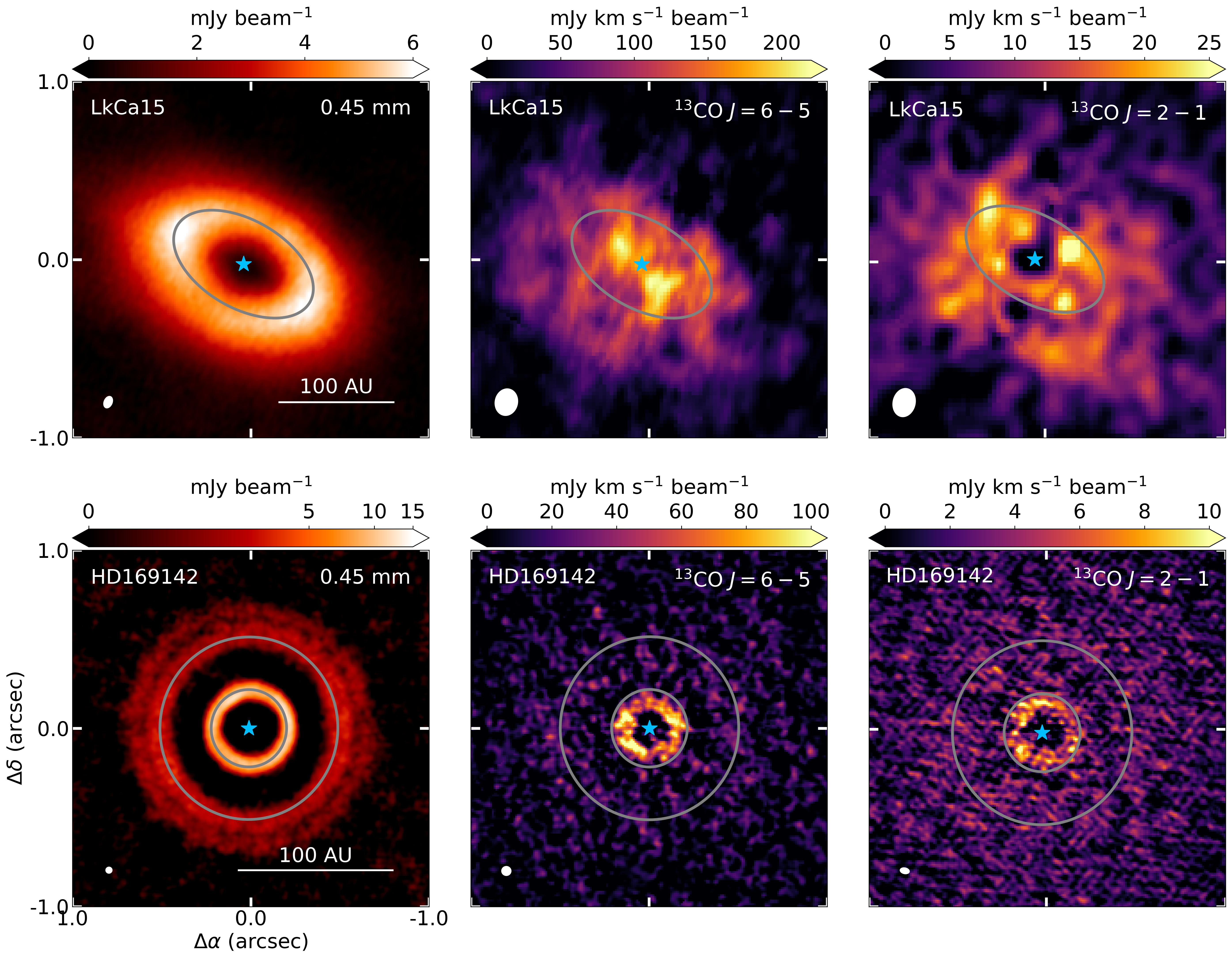}
        \caption{Band~9 continuum (left), moment 0 maps of $^{13}$CO $J=6-5$ (middle) and $^{13}$CO $J=2-1$ (right) after Keplerian masking in the LkCa15 (top) and HD~169142 (bottom) disks. An asinh stretch is applied to the 0.45~mm continuum for HD~169142 to highlight the outer ring. Note the difference in the emission morphologies for the different panels, especially between the gas and dust. The position of the star is marked with a blue star in the center and the rings in the continuum are indicated with grey ellipses in each panel. The beam is indicated by the white ellipse on the bottom left corner of each panel and a 100~AU scalebar is indicated in the bottom right corner. }
        \label{fig:mom0s}
\end{figure*}

In this work we use new ALMA observations of the $^{13}$CO $J=6-5$ transition in the LkCa15 and HD~169142 disks (2017.1.00727.S; PI: J. Szulágyi) as well as archival ALMA observations of the CO, $^{13}$CO, and C$^{18}$O $J=2-1$ lines, see Table~\ref{tab:obs}. The continuum images are used to identify the dust cavity in both disks. All CO transitions for LkCa15 are consistent with a source velocity of 6.3~km~s$^{-1}$. For HD~169142, a source velocity of 6.9~km~s$^{-1}$ is found. The line ratio of the $^{13}$CO $J=6-5$ to the $J=2-1$ transition is used to constrain the temperature across the cavities of both disks, and the optical depth of the $J=2-1$ transition is determined from the ratio with the C$^{18}$O $J=2-1$ transition. Finally, we also use the brightness temperature of optically thick lines as a temperature probe. Subtracting the continuum may remove line flux from the CO isotopologues as the continuum and the molecular lines may be optically thick in these disks. In this case, the continuum flux at the frequencies of the CO isotopologue emission is overestimated as it is absorbed by the CO isotopologues \citep{Isella2016, Weaver2018}. Hence, we do not subtract the continuum for the calculation of the peak brightness temperatures, but we do subtract it for the ratios of not very optically thick lines.

The spatial resolutions of the data for LkCa15 range from $\sim0\farcs07$ to $0\farcs36$. The former resolves the dust cavity with $\sim12$ beams along the major axis of the bright ring in the Band~9 continuum, but the latter spatial resolution is comparable to the minor axis of the dust cavity causing flux of the outer disk to be smoothed into the dust cavity. All data in the HD~169142 disk have a high spatial resolution of $\sim0\farcs05$ used in this work, resolving the dust cavity fully with $\sim9$ beams within the first ring in the Band~9 continuum. 
The Band~9 data for LkCa15 are taken in two different ALMA configurations. The first one has a very high spatial resolution of $\sim0\farcs05$ and a small maximum recoverable scale (MRS) of $0\farcs6$, and the second one has an intermediate spatial resolution of $\sim0\farcs3$ and a larger MRS of $2\farcs1$. The latter data were self-calibrated using two rounds of phase calibration and one round of phase and amplitude calibration. We refer to these data as the intermediate resolution data. These self-calibrated data were also combined with the high resolution data to create a combined high resolution image with a large MRS. We refer to this as the combined high resolution image. Following the works of \citet{Czekala2021} and \citet{Oberg2021}, we applied the JvM correction ($\epsilon=0.20$ for the $^{13}$CO $J=6-5$ transition, and $\epsilon=0.23$ for the continuum) to the residuals to correct for the difference between the dirty and clean beams. Other datasets where both high and intermediate spatial resolution data are available were not combined, instead they were imaged separately. None of the other images are self-calibrated, but all images are corrected for the primary beam response.

\begin{figure*}
        \centering
             \includegraphics[width=\textwidth]{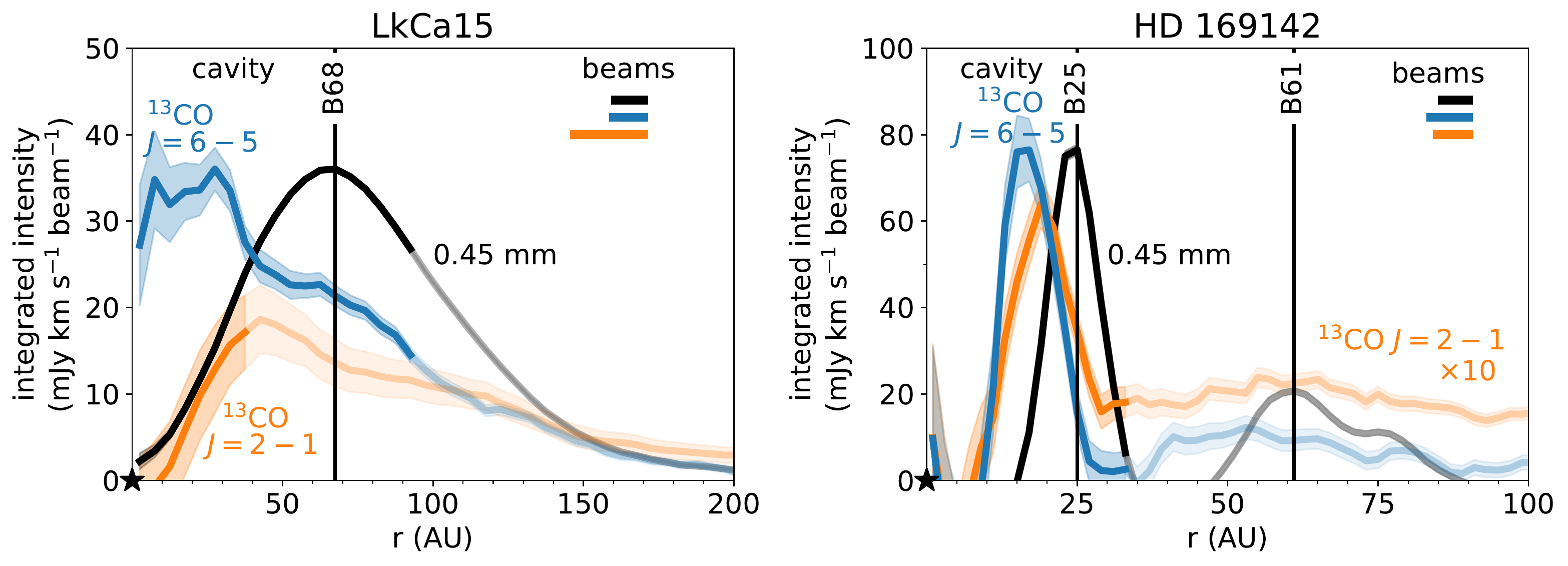}
        \caption{Deprojected azimuthally averaged radial profiles of the $^{13}$CO $J=6-5$ and $J=2-1$ transitions and the scaled $0.45~$mm continuum in the LkCa15 (left) and HD~169142 disks (right). Left: note that $^{13}$CO $J=6-5$ line intensity is roughly constant in the inner $\sim30$~AU in the LkCa15 disk, whereas the 0.45~mm continuum shows a dust cavity. The intensity outside the MRS of the observations is shown in a lighter shade of the corresponding colors. The position of the star is indicated by the black star in the bottom left corner and the beam sizes are indicated by horizontal bars in corresponding colors. Furthermore, the rings in the dust are indicated with the vertical black lines. Note that the MRS of the $^{13}$CO $J=2-1$ line is small compared to the beam size due to the $uv$-taper applied to this dataset.}
        \label{fig:aziavgmom0}
\end{figure*}

For HD~169142, only the high spatial resolution scheduling block was observed. Similar to the intermediate spatial resolution data for LkCa15, we self-calibrated the data covering the $^{13}$CO $J=6-5$ transition using two rounds of phase calibration and one round of phase and amplitude calibration. For the continuum, we used the product data from the ALMA archive as these data were taken during excellent weather conditions and the maximum possible bandwidth was used for the product data, leading to a very good signal-to-noise ratio.

To constrain the temperature in the cavities of both disks, we not only need the $^{13}$CO $J=6-5$ transition, but also a lower $J$ transition, as this provides a good lever arm in upper energy levels. Both disks have suitable archival ALMA observations of the $J=2-1$ transition of CO isotopologues. Therefore, we compare the $^{13}$CO $J=6-5$ transition ($E_{\mathrm{up}} = 111.1$~K) to the $J=2-1$ transition ($E_{\mathrm{up}} = 15.9$~K). For HD~169142, we reimaged the data presented in \citet{Perez2019} and for LkCa15, we use two different datasets. One covers $^{12}$CO at a moderate spatial resolution of $\sim0\farcs35$, and the other one covers $^{13}$CO and C$^{18}$O at moderate ($\sim0\farcs35$) and high spatial resolution ($\sim0\farcs15$). The high spatial resolution of the latter dataset lowers the signal-to-noise ratio greatly compared to the intermediate spatial resolution images. Therefore, the high resolution $^{13}$CO $J=2-1$ transition is imaged using natural weighting and an $0\farcs15$ $uv$-taper. This enhances the signal-to-noise ratio while still resolving the gas cavity.

All observations were imaged using \texttt{multiscale} in the \texttt{tclean} function within CASA version 5.4.0. (combined image of $^{13}$CO $J=6-5$ for LkCa15), 5.6.0 (self-calibration) or 5.7.0 (all lines except the combined image of $^{13}$CO $J=6-5$ for LkCa15) \citep{McMullin2007, Cornwell2008}. For each image, we chose the robust parameter that provides the best balance between the signal-to-noise ratio and the angular resolution. In order to compare the gas cavity with the outer disk, it is crucial to resolve the inner gas cavity. Therefore, we favoured high angular resolution over a high signal-to-noise ratio. Each cube is imaged both before and after continuum subtraction, where the former is used to find the brightness temperature and the latter was used for the other data products.

A Keplerian mask\footnote{\url{https://github.com/richteague/keplerian_mask}} was used to create a mask for the data while cleaning. The robust parameters, resulting disk integrated fluxes, beam sizes, sensitivities, and maximum recoverable scales can be found in Table~\ref{tab:obs} for the observations in the LkCa15 and HD~169142 disks. The integrated flux of the high resolution $^{13}$CO $J=2-1$ transition in LkCa15 is a factor of 5 smaller than the intermediate resolution counterpart because most of the flux at scales larger than $0\farcs5$ (79~AU diameter) is resolved out in the high resolution dataset. The integrated fluxes of the $^{12}$CO, $^{13}$CO, and C$^{18}$O $J=2-1$ transitions for HD~169142 are 34-49\% lower than those reported in \citet{Fedele2017}, due to the smaller MRS of the data used in our work. The shortest baseline of the $J=2-1$ data used in this work is 41.1~m, which is identical to the shortest baseline of the dataset covering the $^{13}$CO $J=6-5$ transition in the HD~169142 disk. Furthermore, the MRS of our two datasets are both $0\farcs6$ (35~AU radius), therefore the $J=6-5$ and $J=2-1$ datasets are expected to have a similar sensitivity to larger scales. 
The MRSs of the data used in this work are typically smaller than the size of the LkCa15 and HD~169142 disks. Therefore, the brightness temperatures may be underestimated at radii similar to the MRSs. The line ratios may also be affected by the MRSs causing small differences e.g. between the $^{13}$CO $J=6-5$ to $J=2-1$ line ratio at (combined) high spatial resolution and that at intermediate spatial resolution in the LkCa15 disk. This is because the MRS of the combined high resolution $^{13}$CO $J=6-5$ image is larger than that of the high resolution $J=2-1$, whereas the opposite is the case for the intermediate resolution data. Therefore, the $^{13}$CO $J=6-5$ to $J=2-1$ line ratio of the (combined) high resolution will be slightly higher than that of the intermediate resolution data.

\begin{figure*}
        \centering
             \includegraphics[width=\textwidth]{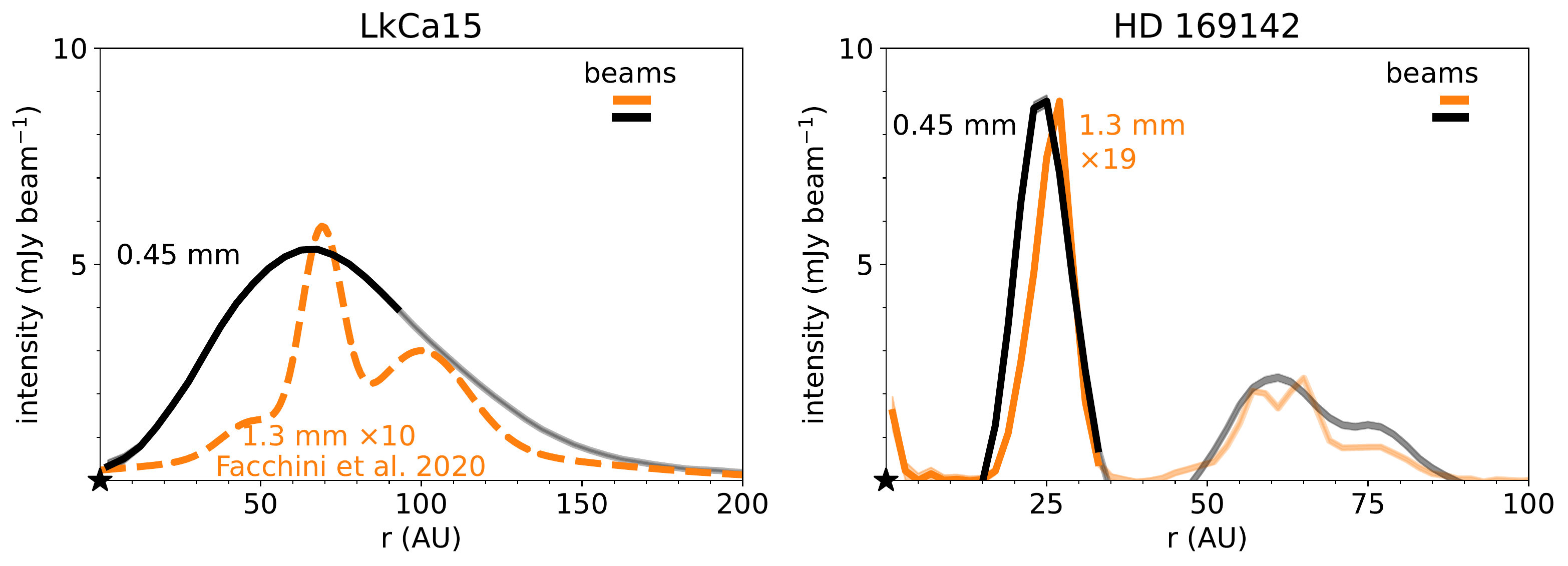}
        \caption{Deprojected azimuthally averaged profiles of the (combined) high resolution Band~9 continuum (black) in the LkCa15 (left) and HD~169142 disks (right). The intensity outside the MRS of the observations is shown in a lighter shade of the corresponding colors. For comparison the fit to the LkCa15 Band~6 continuum by \citet{Facchini2020} (multiplied by 10 for the ease of comparison) and the observed HD~169142 Band~6 continuum (multiplied by 19 to match the Band~9 absolute intensity) are shown in orange.
         The beams are indicated in corresponding colors in the top right corner and the position of the star is marked with the black star in the bottom left corner. Note that the Band~6 continuum in the LkCa15 disk is highly structured whereas the Band~9 continuum shows a single ring. }
        \label{fig:aziavgcont}
\end{figure*}

Finally, the data were imaged to specific beam sizes to compare the $^{13}$CO $J=6-5$ data to the corresponding $^{13}$CO $J=2-1$ transition. For LkCa15, a common restoring beam parameter of $0\farcs36\times 0\farcs30\ (26.6 \degree)$ is used to compare the azimuthally averaged radial profiles of different transitions and isotopologues. This procedure increases the beam size by at most $0\farcs05$. Similarly, a $0\farcs057\times 0\farcs054\ (88.6\degree)$ restoring beam was used for the Band~6 and Band~9 data for HD~169142. As the combined high resolution $^{13}$CO $J=6-5$ data for LkCa15 is a combined data product, these were not convolved using the restoring beam parameter but an $uv$-taper was applied. This resulted in a cube with a $0\farcs15\times 0\farcs13\ (-12.2\degree)$ beam, which was then smoothed using the \texttt{imsmooth} function in casa to match the $0\farcs17 \times\ 0\farcs13 (-13.7\degree)$ beam of the high resolution $^{13}$CO $J=2-1$ data. 

To improve the signal to noise in the moment 0 and peak intensity maps and on the azimuthally averaged radial profiles we used the same Keplerian mask that was also used during the cleaning process to mask any pixels that are not expected to contribute based on the Keplerian rotation of the disk. The noise on the azimuthally averaged radial profile is calculated from the channel rms using error propagation. First the noise on the moment 0 map, $\sigma_{\mathrm{mom0}}$ (in mJy~km~s$^{-1}$~beam$^{-1}$), is calculated using the noise in one channel, $\sigma_{\mathrm{chan}}$ (in mJy~beam$^{-1}$), the number of channels included for each pixel in the moment 0 map, $N_{\mathrm{chan}}$, and the velocity resolution $\Delta V$ (in km~s$^{-1}$):
\begin{align}
\sigma_{\mathrm{mom0}} &= \sigma_{\mathrm{chan}}\sqrt{N_{\mathrm{chan}}}\Delta V.
\end{align}
Then we average this over an annulus and correct for the number of independent measurements using the beam size:
\begin{align}
\sigma_{\mathrm{azimuthal\ average}} &= \sqrt{\frac{1}{N_{\mathrm{beams}}} \sum_{\mathrm{pix}} \sigma_{\mathrm{mom0}}^2},
\end{align}
where $N_{\mathrm{beams}}$ is the number of independent beams per annulus, restricted to be $\geq 1$. 
The 10~\% absolute calibration error is not included in these estimates, but it is included in the error on the total flux. 

In summary, both the Band~9 and Band~6 LkCa15 data have spatial resolutions of $\sim0\farcs07-0\farcs36$, where the latter is similar to the size of the dust cavity. The potential effects of this are discussed in Section~\ref{sec:results_lk}. The data for HD~169142 have a much higher angular resolution of $\sim0\farcs05$. Therefore, the dust cavity in the HD~169142 disk is fully resolved with $\sim9$ beams inside the first ring in the Band~9 continuum data and 5 beams inside the ring in the $^{13}$CO $J=6-5$ data.

\section{Results}\label{sec:results}

The (combined) high resolution Band~9 continuum and the $^{13}$CO $J=6-5$ and $J=2-1$ Keplerian masked integrated intensity maps are presented in Fig.~\ref{fig:mom0s}. 
The Band~9 continuum data show a very clear dust cavity in both disks. This is in contrast to the $^{13}$CO $J=6-5$ and $J=2-1$ transitions that clearly peak inside the cavity seen in the dust, as shown in Fig.~\ref{fig:aziavgmom0}.  Channel maps of the $^{13}$CO $6-5$ data are presented in Figs.~\ref{fig:chanmaps}, \ref{fig:chanmapsHD} in Appendix~\ref{app:obs} and the deprojected azimuthally averaged radial profiles of the $^{12}$CO and C$^{18}$O $J=2-1$ transitions in Fig.~\ref{fig:aziavgmom0app}.

\begin{figure*}
        \centering
             \includegraphics[width=\textwidth]{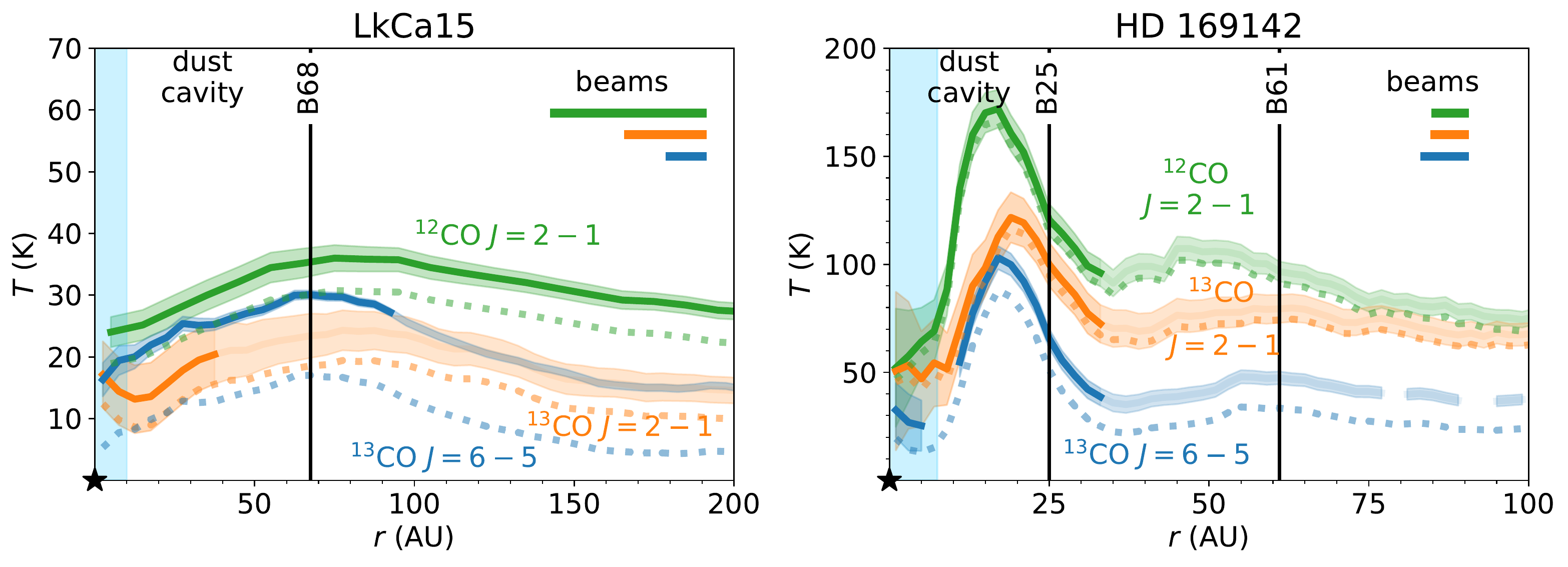}
        \caption{Deprojected azimuthally averaged radial profiles of brightness temperature of the $^{12}$CO $J=2-1$ (green), $^{13}$CO $J=2-1$ (orange), and $^{13}$CO $J=6-5$ (blue) transition for LkCa15 (left) and HD~169142 (right). The solid lines show the brightness temperature calculated using the Planck function and the faded dotted colors show the Rayleigh-Jeans approximation. The vertical black lines indicate the rings seen in the Band~9 continuum and the light blue region starting from 0~AU indicates the gas cavity. The MRS of the HD~169142 $J=6-5$ data corresponds to a radius of 35~AU.}
        \label{fig:aziavgTb}
   
\end{figure*}

\subsection{Dust}

The Band~9 continuum is compared to the Band~6 continuum in both disks in Fig.~\ref{fig:aziavgcont}. The Band~6 continuum in the LkCa15 disk (left hand panel) is the fit from \citet{Facchini2020} and clearly shows two narrow bright rings at 69 and 100~AU and a weaker ring at 47~AU. The Band~9 data on the other hand only show a broad single ring at 68~AU, regardless of the similar beam compared to the Band~6 data, suggesting that the Band~9 continuum is highly optically thick, hiding variations in the column density. Alternatively, the larger grains could be trapped in pressure traps whereas the somewhat smaller grains are not (e.g. \citealt{Pinilla2016}).

In contrast, the morphologies of the very high resolution 0.45~mm and 1.3~mm continuum for HD~169142 are very similar and only differ by a constant factor of 19 in intensity across the disk (right hand panel Fig.~\ref{fig:aziavgcont}). This factor of 19 in intensity is 2.3 times larger than what is expected for optically thick emission. The great similarity in morphology indicates that the trapping mechanism is more efficient in the HD~169142 disk than in the LkCa15 disk.   
Both the Band~6 and Band~9 continuum observations in the HD~169142 disk peak at very similar radii of 27~AU and 25~AU, respectively. The three rings seen at 57, 64, and 76~AU in the high resolution Band~6 data \citep{Perez2019} are observed as two marginally separated rings at 59~AU and 75~AU in the Band~9 data. The fact that we do not see three individual rings is likely due to the slightly larger beam size of our dataset. Finally we note that the high resolution data have some emission with negative intensity between the bright rings. This is likely due to the very high angular resolution, which results in a maximum recoverable scale smaller than the disk. The continuum data will be analysed in more detail in a future paper.

\subsection{Gas}
\subsubsection{Radial profiles}

The $^{13}$CO $J=6-5$ and $J=2-1$ transitions in the LkCa15 disk (top row in Fig.~\ref{fig:mom0s} and left panel in Fig.~\ref{fig:aziavgmom0}) show very different morphologies compared to the 0.45~mm continuum and to each other. The $^{13}$CO $J=6-5$ data have a roughly constant integrated intensity in the inner $\sim30$~AU, whereas the continuum and the $J=2-1$ data both show a clear cavity. The $^{13}$CO $J=2-1$ gas cavity is marginally resolved at our spatial resolution. These data suggest that there is still some gas in the dust cavity. Furthermore, they show that the line ratio of the $^{13}$CO $J=6-5$ to the $J=2-1$ line  increases steeply inside the dust cavity, highlighting that molecular excitation, controlled by the gas temperature, and the gas surface density play a role in the molecular line emission.

The C$^{18}$O emission for LkCa15 decreases with decreasing radius between 50 and 100~AU but there appears to be some emission inwards of 50~AU (see Fig.~\ref{fig:aziavgmom0app}). This is likely due to the low signal-to-noise ratio and due to beam dilution of the gas cavity as the major axis of the C$^{18}$O beam is comparable to the minor axis of the dust cavity. Accordingly, C$^{18}$O emission from the region just outside the dust cavity is smoothed into the cavity, which affects our further analysis in Section~\ref{sec:results_lk}. Therefore, we proceed to treat this as a $3\sigma$ upper limit on the C$^{18}$O emission inside the dust cavity.

In the case of the HD~169142 disk, the $^{13}$CO $J=6-5$ and $J=2-1$ lines peak $\sim10$~AU inside the dust ring and just inside the ring seen in scattered light \citep{Tschudi2021}. Both $^{13}$CO lines show a clear gas cavity inwards of 10~AU, similar to the structure seen in other transition disks like DoAr44, IRS~48, HD~135344B, HD~142527, and PDS~70 \citep{Bruderer2014, Perez2015, vanderMarel2015, vanderMarel2016, vanderMarel2018, Facchini2021}. Finally, the $^{13}$CO $J=2-1$ transition is detected out to much larger radii than the $J=6-5$ transition and the continuum data in both disks.

\begin{figure*}
        \centering
        \includegraphics[width=\textwidth]{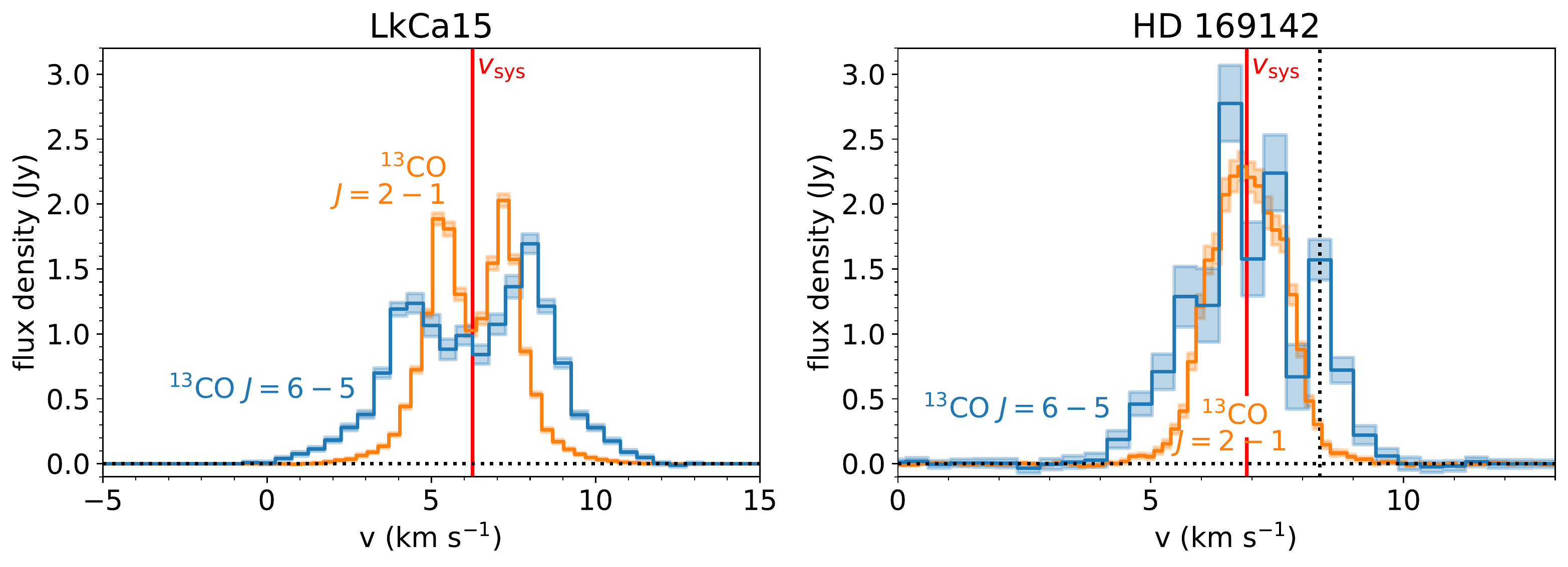}
        \caption{Keplerian masked spectrum of $^{13}$CO $J=6-5$ (blue) and $^{13}$CO $J=2-1$ (orange) of the LkCa15 (left) and HD~169142 disk (right). The shaded area indicates the $1\sigma$ uncertainty on the spectra. The vertical black dotted line in the HD~169142 panel indicates the bump in the spectrum corresponding to the inner ring at 17~AU seen in the azimuthally averaged radial profile of the $^{13}$CO $J=6-5$ transition. Note that for S/N-ratio considerations, we show the $^{13}$CO $J=2-1$ data at intermediate spatial resolution for LkCa15. The spectrum of the $^{13}$CO $J=6-5$ transition for LkCa15 is obtained from the combined datasets. } 
        \label{fig:spec}
\end{figure*}

\begin{figure*}
        \centering
             \includegraphics[width=\textwidth]{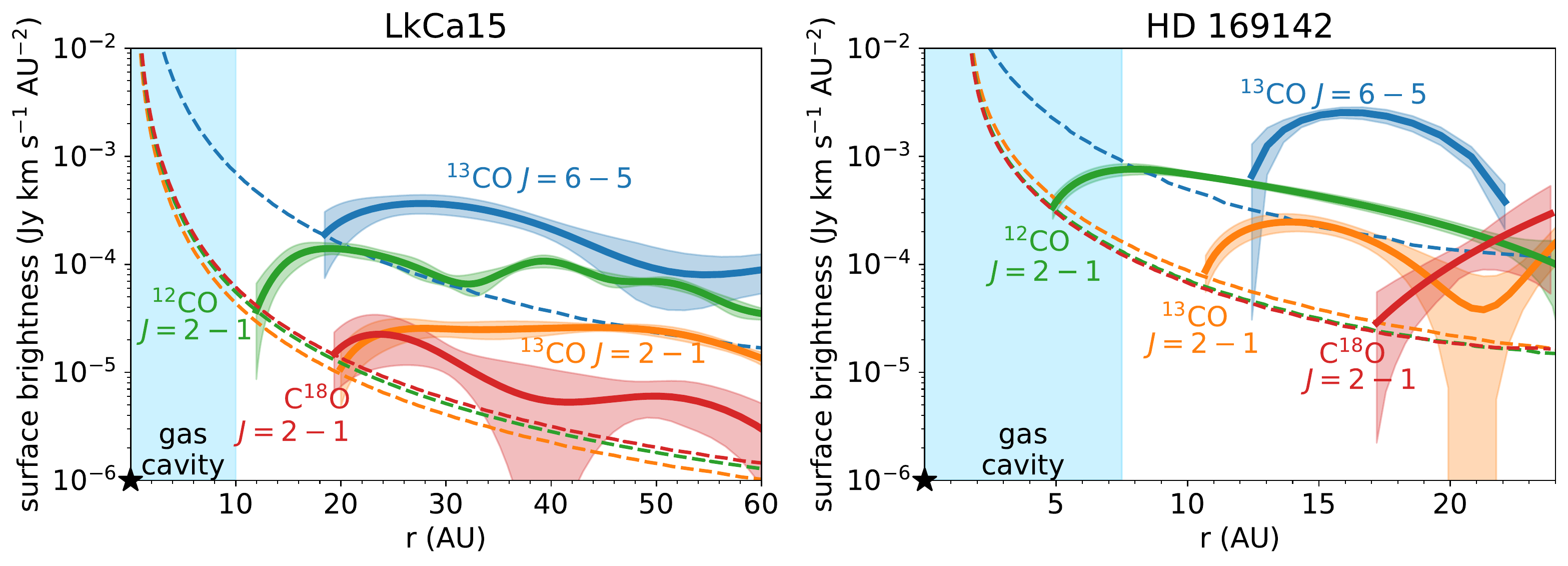}
        \caption{Surface brightness of the $^{13}$CO $J=6-5$ (blue), $^{13}$CO $J=2-1$ (orange), $^{12}$CO $J=2-1$ (green) and C$^{18}$O $J=2-1$ (red) lines inside the dust cavities derived using the kinematics. The dashed lines in corresponding colors indicate the detection limit for each line. }
        \label{fig:kinfitter}
   
\end{figure*}

\subsubsection{Brightness temperatures} \label{sec:Tb}
The brightness temperatures of the two $^{13}$CO and the $^{12}$CO $J=2-1$ transition are presented in Fig.~\ref{fig:aziavgTb}. First, we highlight the difference between the Rayleigh-Jeans approximation (dotted lines) and the full Planck curve (solid lines). The results are different because the Rayleigh-Jeans approximation is only valid if $h\nu \ll kT$, which results in $T \gg 32$~K for the $^{13}$CO $J=6-5$  and $T \gg 11$~K for the $^{12}$CO and $^{13}$CO $J=2-1$ transitions. This criterion is not met for these observations except for the $J=2-1$ data in the HD~169142 disk. Therefore, care should be taken when using the Rayleigh-Jeans approximation for high frequency observations and observations with low peak intensities.

Comparing their brightness temperatures directly shows that the disk around the Herbig star HD~169142 is warmer than that around the T~Tauri star LkCa15. 
The brightness temperature of $^{12}$CO is highest in both disks as this line becomes optically thick highest up in the disk, hence it traces a warmer layer closer to the surface of the disk. The $^{13}$CO $J=2-1$ transition traces a lower and hence colder layer due to its lower optical depth (see Fig.~\ref{fig:tau}). The brightness temperature of both $^{13}$CO $J=6-5$ lines is lower than that of the $^{12}$CO $J=2-1$ lines. This is due to the lower optical depth of the $J=6-5$ transition and due to its lower spectral resolution. The latter lowers the peak flux by 6-13~K (40-60\%) in LkCa15  and 10-15~K (10-30\%) in HD~169142, see Appendix~\ref{app:Tbspec}.
All CO lines may become more optically thin in the outer disk and in the gas cavities, which lowers their brightness temperature. In the case of LkCa15, the gas inside the dust cavity may be similarly warm as in the HD~169142 disk, but this cannot be determined from the brightness temperatures presented here due to beam dilution and, optical depth effects in the inner disk.

For HD~169142, the $^{12}$CO line indicates gas as warm as 170~K. The brightness temperature of the $^{13}$CO $J=6-5$ transition for HD~169142 is somewhat lower than the corresponding $^{13}$CO $J=2-1$ transition, which is not what is expected. The $J=2-1$ emission is optically thick (see also Fig.~\ref{fig:tau}), hence its brightness temperature is tracing the kinetic temperature in the disk layer where its optical depth equals 1. For these temperatures, the molecular excitation is such that the optical depth of the $J=6-5$ line is comparable to or larger than that of the $J=2-1$ line. Therefore, the $J=6-5$ transition should trace a higher layer and warmer temperatures. 
The reason that we do not observe this is likely due to the small maximum recoverable scale of $0\farcs6$ (35~AU radius) of the high resolution data, which resolves out some of the flux inside this radius. Furthermore, the shortest baseline of the $^{13}$CO $J=6-5$ data is longer than that of the $=2-1$, causing the $J=6-5$ data to be less sensitive to larger spatial scales. This is not reflected in the MRS of $0\farcs6$, because the MRS is calculated using the 5$^{th}$ percentile of the baseline distribution and is calculated assuming that the morphology of the emission resembles a uniform disk. Our observations show that the morphology of the $^{13}$CO $J=6-5$ and $J=2-1$ lines is different. Therefore, the low $^{13}$CO $J=6-5$ brightness temperature may be partially due to the small MRS. Finally, we note that the $^{13}$CO $J=6-5$ transition is not spectrally resolved for radii larger than $\sim30$~AU, lowering the peak flux and hence the brightness temperature by 10-15~K ($\sim$10-30\%, see Appendix~\ref{app:Tbspec}).

\subsubsection{Spectra} \label{sec:spectra}

The azimuthally averaged radial profiles discussed in the previous section show a very clear difference between the morphology of the $^{13}$CO $J=6-5$ transition and the $J=2-1$ transition for LkCa15. The spectra extracted from the Keplerian masked images of these two transitions are presented in Fig.~\ref{fig:spec}.

Both disks clearly show that the $^{13}$CO $J=6-5$ the line profile is broader than that of the $J=2-1$ line. This is another indication that the $J=6-5$ transition is brighter than the $J=2-1$ transition in the inner disk, as the innermost regions of the disk are rotating at the highest Keplerian velocity:
\begin{align}
R = GM_{\star}\left (\frac{\sin(i)}{V_{\mathrm{Kep}}}\right )^2,
\end{align}
with $G$ the gravitational constant, $M_{\star}$ the mass of the star, $i$ the disk inclination and $V_{\mathrm{Kep}}$ is the Keplerian velocity of the gas.

\citet{Bosman2021} used this principle to derive the surface brightness of molecular line emission in the innermost disk regions with a spatial resolution smaller than the beam. Using this kinematics fitting program, we are able to reconstruct the surface brightness of the molecular line emission in the innermost regions of the LkCa15 and HD~169142 disks. 
A $1\farcs0$ region was used to extract the spectrum for LkCa15 and a $0\farcs5$ region for HD~169142. No Keplerian mask was applied. Furthermore, the spectral resolution of the model was set to the native spectral resolution of the data and the spectra were modelled out to $\pm$60~km~s$^{-1}$ from the velocity corresponding to the peak flux density. The surface brightness is fitted up to 80~AU (LkCa15) and 25~AU (HD~169142). Finally, $\geq 2$ (LkCa15) and $\geq 3$ (HD~169142) velocity resolution elements per model fitting point were used for the surface brightness. The model fitting points are separated by $\geq6$~AU (LkCa15) and $\geq 2$~AU (HD~169142), effectively increasing the signal-to-noise ratio in the LkCa15 disk.

The resulting radial profiles for the $^{12}$CO, $^{13}$CO, and C$^{18}$O $J=2-1$ and the $^{13}$CO $J=6-5$ lines are presented in Fig.~\ref{fig:kinfitter}. The dashed lines indicate the detection limit for each radial profile in corresponding colors. In the LkCa15 disk, all transitions except for $^{12}$CO $J=2-1$ drop below their detection limit around 20~AU. This is consistent with the azimuthal averages presented in Fig.~\ref{fig:aziavgmom0} but now resolving the emission to a smaller spatial scale than what is obtained with the traditional azimuthally averaged radial profiles presented in Fig.~\ref{fig:aziavgmom0app}. The $^{12}$CO $J=2-1$ line is detected down to 12~AU, as $^{12}$CO has the highest column density and hence the highest signal-to-noise ratio in the line wings, making it easier to detect than the less abundant isotopologues. 
As no steep increase in the surface brightness in the inner disk is detected in any of the lines presented here and the turnover of each line at $10-20$~AU is significant compared to the corresponding detection limit, the results from the kinematically derived radial profiles can be interpreted as a $\sim10$~AU cavity in the gas in the LkCa15 disk. Notably, this is significantly smaller than the cavity seen in the dust that peaks at 68~AU. 

In the HD~169142 disk, the $^{13}$CO $J=6-5$ and $J=2-1$ transitions both drop below the detection limit at $\sim10-12$~AU, similar to what is seen in the traditional azimuthally averaged radial profiles. The C$^{18}$O $J=2-1$ emission is only detected from 17~AU outward due to its low signal-to-noise ratio. On the other hand, $^{12}$CO $J=2-1$ is detected down to the smallest radius of 5~AU, similar to LkCa15. This is supported by the signal-to-noise ratio in the line wings (right hand panel Fig.~\ref{fig:kinfitterspec}), where the $^{12}$CO $J=2-1$ transition is detected out to the largest velocities. As all lines quickly drop below their detection limits and their turn overs are significant, the results from the kinematically derived radial profiles in the HD~169142 disk can be understood as a deep gas cavity at $\sim5-10$~AU.

In summary, the derived gas cavity radii in the LkCa15 and HD~169142 disks are smaller by $\sim15-60$~AU than their dust cavity radii, consistent with models of planet-disk interaction  and formation of dust traps (e.g., \citealt{Pinilla2012, deJuanOvelar2013}).  
Furthermore, the brightness temperature indicates that the gas in the dust cavity in the HD~169142 disk is warmer than that in the outer disk. This is not seen in the brightness temperature in the LkCa15 disk due to beam dilution that lowers the peak intensity inside the dust cavity. The gas temperature inside these regions will be further constrained in the next section.

\section{Analysis} \label{sec:ana}

To analyse the data in terms of physical quantities, we first briefly reiterate the relevant equations to derive the temperature, column density and estimates of the optical depth using CO line ratios and assuming that all levels are characterised by a single excitation temperature, i.e., the common assumption that the lines are emitting from the same layer in the disk. Next, we apply this to the observations to investigate the differences between the cavities and the outer disks of LkCa15 and HD~169142.

\subsection{Temperature and column density determination}

\begin{figure}
   \centering
           \begin{subfigure}[b]{0.475\textwidth}
            \centering
            \includegraphics[width=\textwidth]{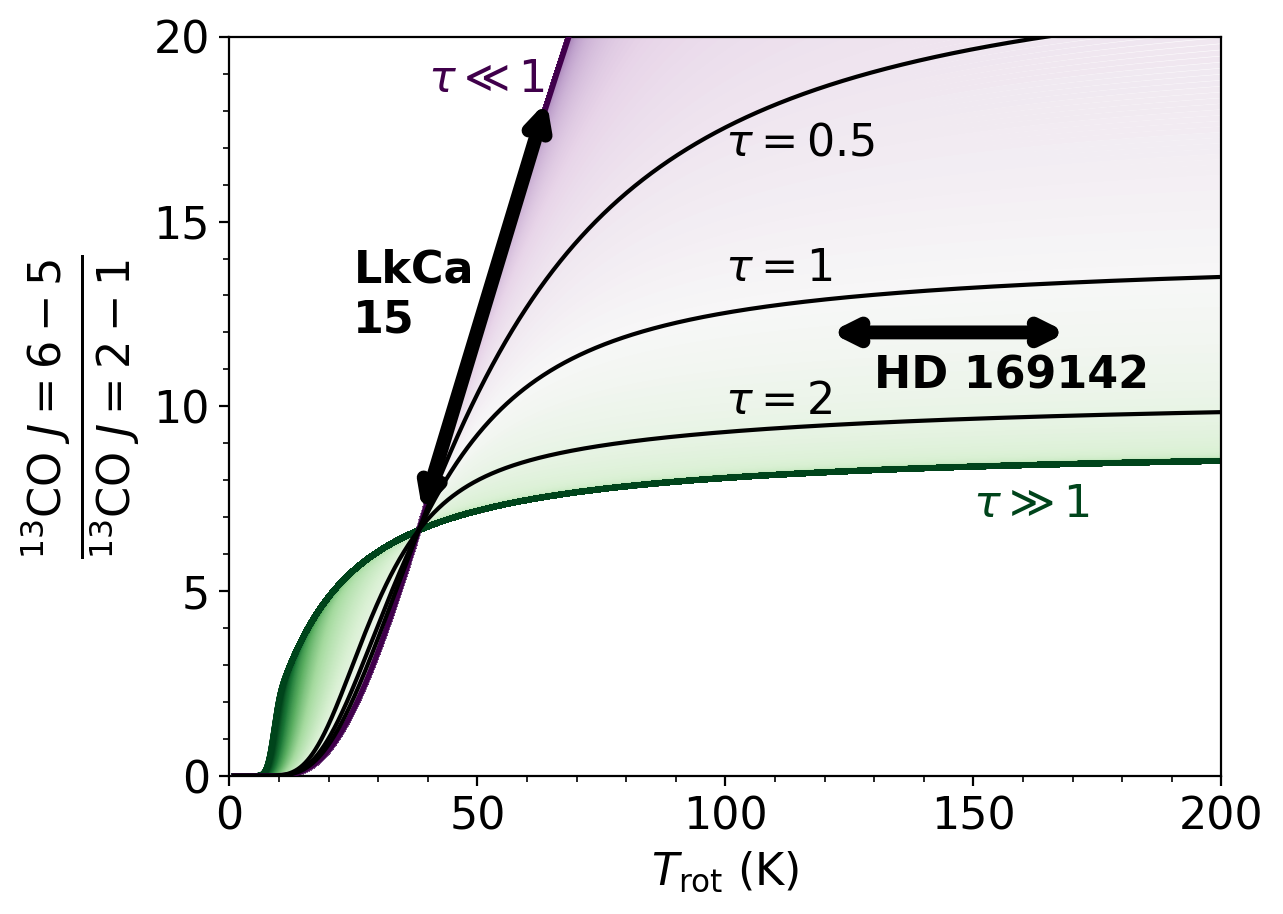}
        \end{subfigure}
   \caption{Model line ratio of $^{13}$CO $J=6-5$ to $^{13}$CO $J=2-1$ as function of rotational temperature for very low (purple) to very high (green) optical depths of $^{13}$CO 2-1. The contours highlight the line ratio for a $^{13}$CO $J=2-1$ optical depth of 0.5, 1 and 2 (black). Note that the line ratio scales linearly with temperature if the emission is optically thin and $25~\mathrm{K} \lesssim T_{\mathrm{rot}} \lesssim 150$~K and that it becomes independent of temperature at a value of 9 if the emission is optically thick and $T_{\mathrm{rot}} \gtrsim 40$~K. The typical values for the line ratio and rotational temperature (LkCa15) and brightness temperature (HD~169142) in their dust cavities are indicated in the figure. }
   \label{fig:lineratioana}
\end{figure}

\begin{figure*}
\centering
        \includegraphics[width=\textwidth]{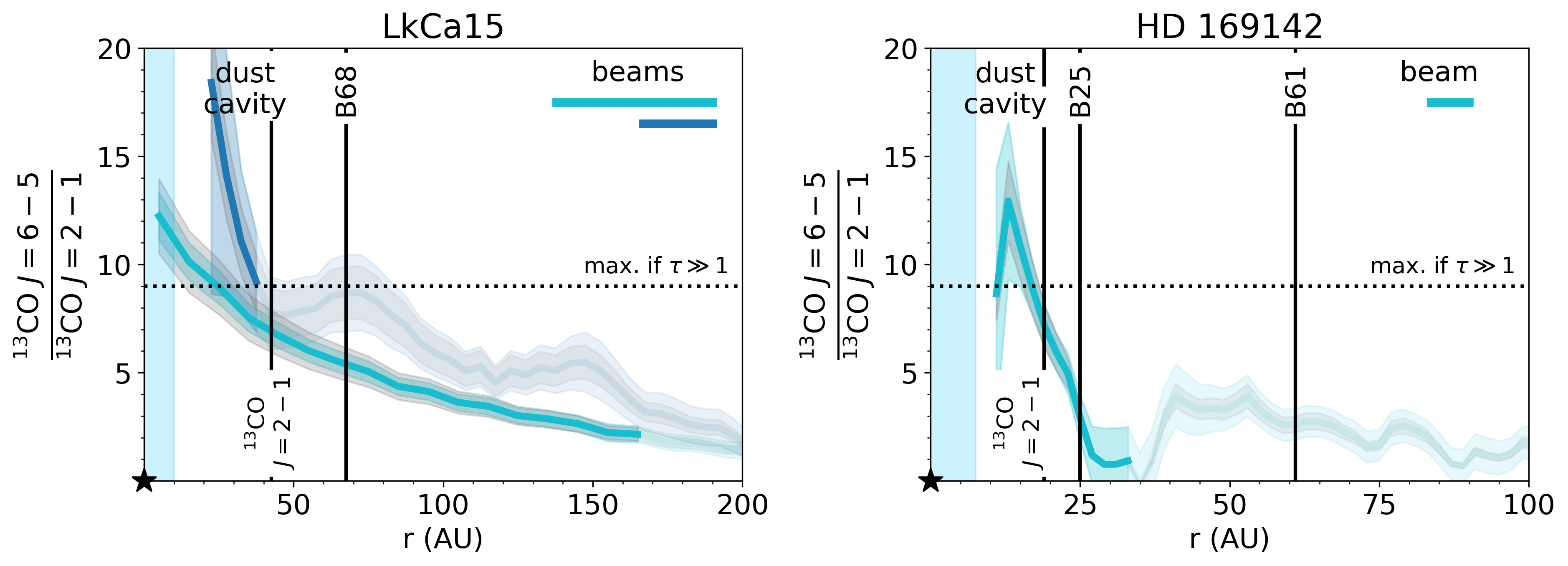}
\caption{Radially resolved $^{13}$CO $J=6-5$ to $^{13}$CO $J=2-1$ line ratio in the LkCa15 (left; (combined) high and intermediate resolution data) and HD~169142 disk (right). The rings in the continuum and the radius where the $^{13}$CO $J=2-1$ transition peaks are indicated with the vertical black lines. The horizontal black dotted line indicates the theoretical maximum line ratio if both lines are optically thick and emit from the same disk region. The beam is indicated with the horizontal bar in the top right corner. The grey shaded region shows the $10~\%$ absolute calibration error. The line ratio is not shown for radii where the $^{13}$CO $J=6-5$ or the $J=2-1$ integrated intensity is below $1.5\sigma$. Furthermore, the intensity outside the MRS of the observations is shown in a lighter shade of the corresponding colors.}
\label{fig:LkCa15ratio6521}
\end{figure*}

  \begin{figure*}
        \centering
        \includegraphics[width=\textwidth]{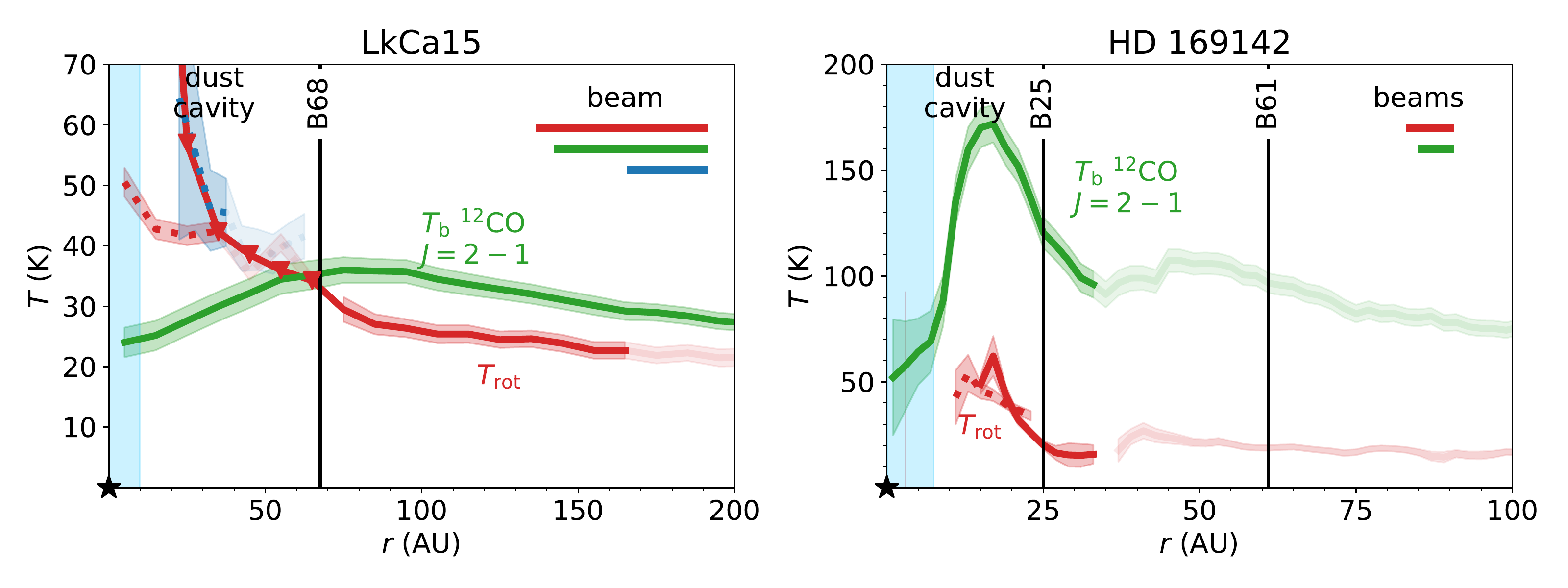}
        \caption{Temperature derived from the line ratio analysis. Rotational temperature (red) derived from the $^{13}$CO $J=6-5$ to $J=2-1$ line ratio including optical depth constraints (solid lines) for LkCa15 (left) and HD~169142 (right). The dotted lines in corresponding colors indicate the rotational temperature inside the dust cavity if the emission were optically thin. For comparison the brightness temperature of the optically thick $^{12}$CO $J=2-1$ transition (green) is shown. Note the difference in the vertical temperature axes. In the case of LkCa15, the blue dotted line shows the rotational temperature assuming optically thin emission for the (combined) high resolution observations. The downward triangles indicate the upper limit on the rotational temperature due to the upper limit on the integrated C$^{18}$O $J=2-1$ intensity in the dust cavity. The intensity outside the MRS of the observations is shown in a lighter shade of the corresponding colors. }
        \label{fig:Trot}
\end{figure*}

  \begin{figure}
        \centering
        \includegraphics[width=\linewidth]{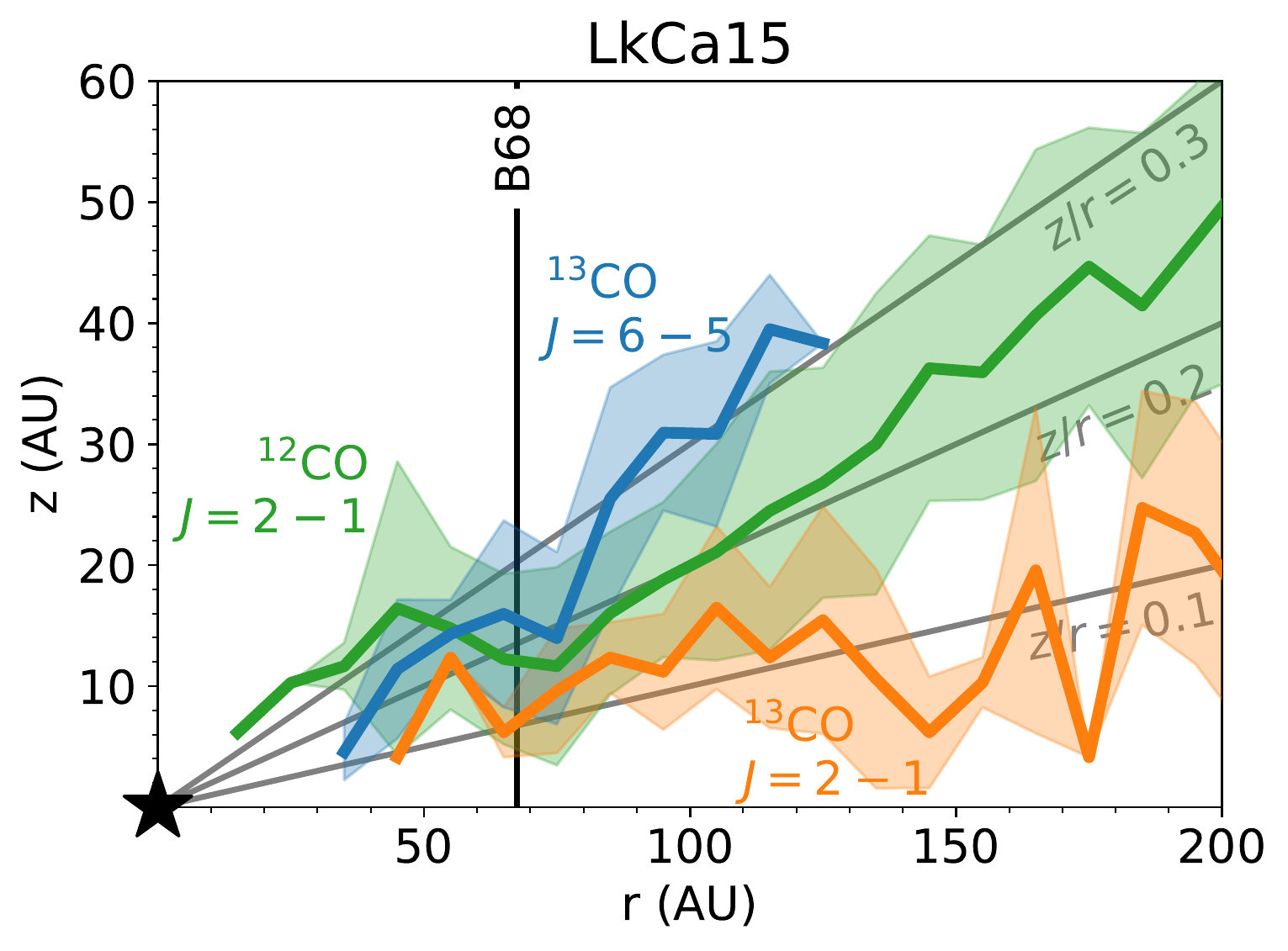}
        \caption{Emitting surface of $^{12}$CO $J=2-1$ (green), $^{13}$CO $J=2-1$ (orange), and $^{13}$CO $J=6-5$ (blue) for LkCa15. The grey lines indicate the emitting surface for $z/r = 0.1, 0.2,$ and $0.3$. }
        \label{fig:emittingheights}
\end{figure}

The temperature structure inside and outside the dust cavity can be found by analysing the $^{13}$CO $6-5$ and $^{13}$CO $2-1$ transitions, as these transitions are far apart in upper energy level ($E_u = 15.9$~K vs $E_u = 111.1$~K). The column density structure can then be derived using the less optically thick C$^{18}$O $2-1$ emission together with the temperature structure from the $^{13}$CO analysis. 

If the emission is optically thin, the column density is proportional to the integrated intensity (e.g., \citealt{Goldsmith1999}):
\begin{align}
N_u^{\mathrm{thin}} = \frac{4\pi F_{\nu}\Delta V}{A_{ul}\Omega hc}, \label{eq:Nthin}
\end{align}
with $N_u^{\mathrm{thin}}$ the column density of the upper level $u$, $F_{\nu}\Delta V$ is the integrated intensity, $A_{ul}$ is the Einstein $A$ coefficient, $\Omega$ is the emitting region, $h$ is the Planck constant, and $c$ is the speed of light. If the emission is optically thick, a correction factor needs to be applied:
\begin{align}
N_u = N_u^{\mathrm{thin}}C_{\tau} \label{eq:Nthick}
\end{align}
where the correction factor is defined as:
\begin{align}
C_{\tau} &\equiv \frac{\tau_{\nu}}{1-e^{-\tau_{\nu}}}.
\end{align}
Note that this expression approaches 1 for $\tau \ll 1$, such that $N_u$ approaches $N_u^{\mathrm{thin}}$. On the other hand, if the emission becomes optically thick, this correction factor scales with $\tau$, whose precise value is uncertain in this regime.

The estimate of the column density of the upper state can also be related to the total column density assuming LTE:
\begin{align}
\frac{N_u}{g_u} &= \frac{N_{\mathrm{tot}}}{Q(T_{\mathrm{rot}})} e^{-E_u/kT_{\mathrm{rot}}}, \label{eqN2g2}
\end{align}
with $Q(T_{\mathrm{rot}})\approx kT_{\mathrm{rot}}/(hB_{\mathrm{0}}) + 1/3$ the partition function of CO at a rotational temperature $T_{\mathrm{rot}}$ with rotational constant $B_{\mathrm{0}}$, and $k$ is the Boltzmann constant. LTE excitation is a valid assumption for $^{13}$CO $J=6-5$ and $J=2-1$ line emission in protoplanetary disks as the critical density is $\sim10^6$~cm$^{-3}-3\times 10^4$~cm$^{-3}$ for the two lines, respectively, which is easily reached even in higher layers. 
This equation can be solved for the temperature (and column density) as we have observations of both $^{13}$CO $J=6-5$ and $J=2-1$.  A detailed description of the methods used to solve this equation and to estimate the optical depth can be found in Appendix~\ref{app:rotdia}.

The result of this analytical analysis is shown in Fig.~\ref{fig:lineratioana}. If the emission is optically thin, the line ratio is linearly proportional to the rotational temperature for $T \gtrsim 25$~K. If, on the other hand, the emission is optically thick, the line ratio is only very weakly dependent on the temperature. The predicted line ratio only increases from 8 to 8.5 between temperatures of 90 to 200~K. For higher temperatures, the line ratio approaches $(\nu_{\mathrm{65}}/\nu_{\mathrm{21}})^2 = 9$, complicating the determination of the temperature.

The line ratio analysis requires that both $^{13}$CO lines emit from the same disk layer. The emitting surfaces of molecular line emission in inclined protoplanetary disks can actually be inferred directly from the channel maps, as emission from an elevated layer appears offset compared to the disk center  \citep{Pinte2018temperature, Law2021vert, PanequeCarrenosubm}. The upper surface in the disk was traced within hand-drawn masks in each channel containing line emission. For each position along the disk major axis, the location where the emission peaks in the near and far side of the disk are converted to an emitting height using their offset compared to the disk center, the Keplerian velocity and the parameters listed in Table~\ref{tab:sources}. Subsequently, the resulting emitting surface was binned in radius to increase the signal-to-noise ratio.

\subsection{Results for LkCa15} \label{sec:results_lk}

The observed line ratio for LkCa15 is presented in Fig.~\ref{fig:LkCa15ratio6521}, where the lines used for the ratio are convolved to the same beam of $0\farcs 36 \times 0\farcs 30\ (26.6\degree)$ or $0\farcs17\times 0\farcs13\ (-13.7\degree)$ in case of the intermediate and (combined) high resolution observations respectively. 
The line ratio of the $^{13}$CO $J=6-5$ to $J=2-1$ transition decreases with radius in the outer disk. This is expected as the temperature of the gas decreases with radius. Furthermore, the data show an increase in the slope inside the dust cavity compared to outside the dust cavity, driven by the drop in $^{13}$CO $J=2-1$ inside 40~AU. Whether this indicates an actual increase in the temperature in the dust cavity or if it is just an optical depth effect is investigated below. The line ratio found from the (combined) high resolution data is consistent with the intermediate resolution data outside 45~AU, within the limitations of the data.
The line ratio of the $^{12}$CO $J=6-5$ transition presented in \citet{vanderMarel2015} and the $^{12}$CO $J=2-1$ transition in a common beam of $0\farcs35\times 0\farcs25\ (-37.5\degree)$ is similar in value and trend with radius to the intermediate resolution line ratio of the $^{13}$CO isotopologue.

The optical depths of the $^{13}$CO $J=6-5$, $J=2-1$ and the C$^{18}$O $J=2-1$ transitions at moderate spatial resolution are derived using Eq.~\ref{eq:tau} and using a $3\sigma$ upper limit on the integrated C$^{18}$O intensity inside the dust cavity ($\leq68$~AU; see Fig.~\ref{fig:tau}) as the radial profile derived from the kinematics clearly suggests a much smaller gas cavity of only $\sim10$~AU. The optical depth $^{13}$CO $J=2-1$ and C$^{18}$O $J=2-1$ transition differ by their isotopologue ratio as this is one of the assumptions used to find the optical depth. The C$^{18}$O $J=2-1$ emission is marginally optically thick in the outer disk with an optical depth between $\sim 0.1$ and $\sim 0.4$ between 70~AU and 200~AU. Consequently, the $^{13}$CO $J=2-1$ emission is optically thick throughout the same disk region. Inside the dust cavity, the optical depths of $\leq2$ for the $^{13}$CO and $\leq0.3$ for the C$^{18}$O $J=2-1$ are upper limits. The optical depth of the $^{13}$CO $J=6-5$ transition follows from the line ratio analysis and lies typically between the optical depth from the aforementioned transitions in the outer disk. Inside the dust cavity, an upper limit on the $^{13}$CO $J=6-5$ optical depth is given. The $^{13}$CO and C$^{18}$O lines could be close to optically thin inside the dust cavity, thus their brightness temperatures are only lower limits to the kinetic temperature.

  \begin{figure*}
        \centering
        \includegraphics[width=\textwidth]{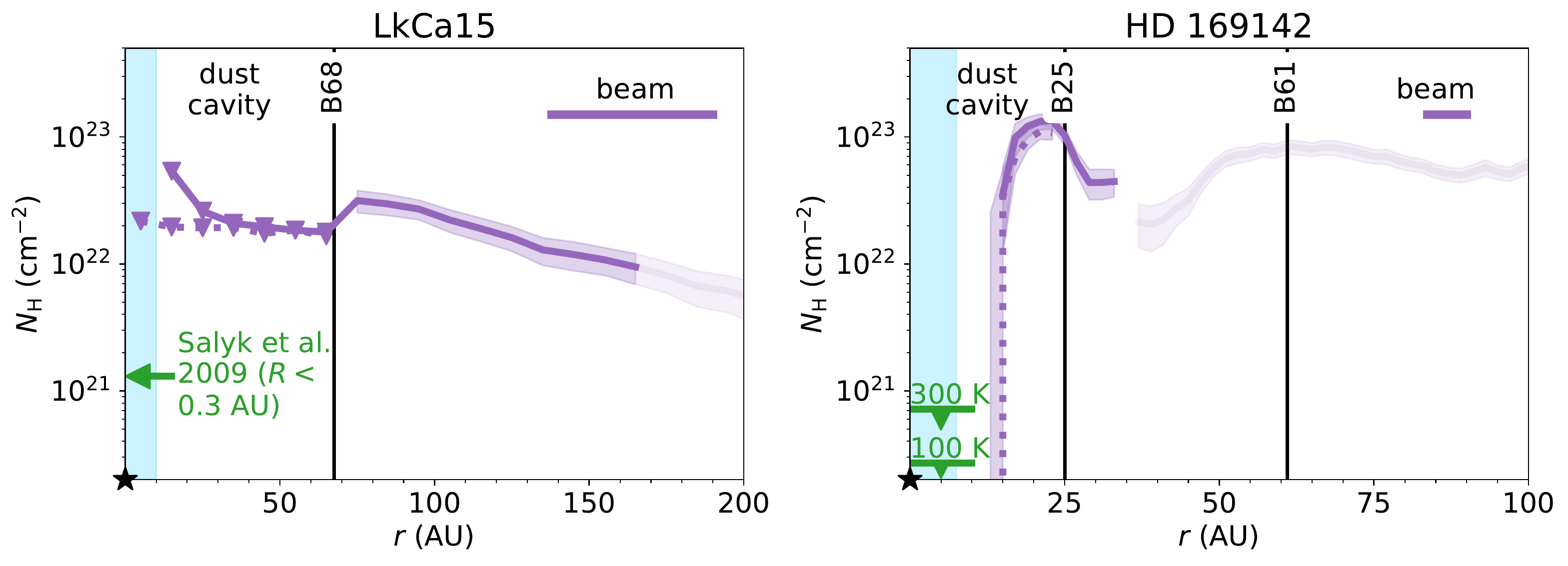}
        \caption{Total hydrogen column density (purple) derived from the $^{13}$CO $J=6-5$ to $J=2-1$ line ratio, the integrated C$^{18}$O $J=2-1$ line intensity and optical depth for LkCa15 (left) and HD~169142 (right). The dotted line indicates the rotational temperature and column density inside the dust cavity if the emission were optically thin. The purple downward triangles in the left panel indicate the upper limit on column density due to the upper limit on the integrated C$^{18}$O $J=2-1$ intensity in the dust cavity. The green bars with downward triangles in the right panel indicate the 3$\sigma$ upper limit on the hydrogen column density derived from the non-detection of $^{12}$CO $J=2-1$ in this region and assuming a temperature of 100~K and 300~K respectively.}
        \label{fig:NHtot}
\end{figure*}

The rotational temperature is presented in the left hand panel of Fig.~\ref{fig:Trot} and increases towards the dust cavity. For comparison, the $^{12}$CO $J=2-1$ brightness temperature is also shown. This transition is surely optically thick outside the dust cavity so the brightness temperature traces the kinetic temperature in the emitting layer of the outer disk. This brightness temperature is $5-10$~K higher than the rotational temperature of $^{13}$CO in the outer disk.
In the inner disk, the line ratio steeply increases towards the star, indicating a steep increase in the gas temperature.

The (combined) high resolution data for LkCa15 do not have an C$^{18}$O $J=2-1$ counterpart. Therefore, the optical depth cannot be determined following the same method as for the intermediate resolution data. Still, the steep increase in the line ratio in the dust cavity and the upper limit on the $^{13}$CO $J=2-1$ optical depth of $\sim2$ in the dust cavity at moderate spatial resolution suggest that the gas in the dust cavity is likely optically thin in the two $^{13}$CO and the C$^{18}$O $J=2-1$ lines. In this case, the temperature derived from the line ratio (blue dotted line in Fig.~\ref{fig:Trot}) is very similar to the rotational diagram analysis on the moderate spatial resolution data.\footnote{The 40~AU radius MRS of the high resolution $^{13}$CO $J=2-1$ line is not expected to affect this result as it covers most of the dust cavity. Therefore, most flux is recovered in this region.}

The hypothesis that the $^{13}$CO $J=6-5$ and $J=2-1$ transitions emit from the same disk region inside the dust cavity is supported by the emitting surfaces presented in Fig.~\ref{fig:emittingheights}. First, we note that the LkCa15 disk is very flat with $z/r\sim0.1$ for the $^{13}$CO $J=2-1$ transition similar to e.g. AS~209, HD~163296 and MWC~480 \citep{Law2021vert}. At radii smaller than 85~AU, the emitting surfaces of these two transitions measured from the (combined) high resolution data are consistent within the errorbars. Moreover, their brightness temperatures are very similar (Fig.~\ref{fig:aziavgTb}), which is as expected if the same layer is traced. Outside this radius of 85~AU, the $^{13}$CO $J=6-5$ and $^{12}$CO $J=2-1$ lines emit from a similar layer at $z/r\sim0.2-0.3$, whereas the $^{13}$CO $J=2-1$ line emits from a lower layer at $z/r\sim0.1$. This is the first time that we derive directly from the data that the $^{13}$CO $J=6-5$ line indeed emits from a higher layer than the $^{13}$CO $J=2-1$ line.

Finally, the total hydrogen column density in the LkCa15 disk is derived using the rotational temperature and the absolute integrated C$^{18}$O intensity at intermediate resolution, assuming a CO abundance of $10^{-4}$ w.r.t. the total number of hydrogen atoms. As the C$^{18}$O emission is marginally optically thick throughout the outer disk, the optical depth of the line is explicitly taken into account. The total column density, presented in Fig.~\ref{fig:NHtot}, increases towards the central star up to $\sim75$~AU. However, the exact radius where the gas column density peaks is smaller as argued in Sect.~\ref{sec:spectra}. Inside the dust cavity, the C$^{18}$O upper limit gives an upper limit of $5\times 10^{22}$~cm$^{-2}$ at 15~AU, consistent with the value of $1.3\times 10^{21}$~cm$^{-2}$ in the inner 0.3~AU using the CO column density derived by \citet{Salyk2009}, and with the modelling result found by \citet{vanderMarel2015}. Furthermore, the derived drop in gas column density inside the dust cavity is a factor of $\geq 2$, consistent with the modelling by \citet{vanderMarel2015}. Finally, we note that the total hydrogen column density in the LkCa15 disk may be higher than the values derived here depending on whether CO is transformed to other species resulting in a CO abundance lower than the assumed value of $10^{-4}$ (Sturm et al. in prep.).

\subsection{Results for HD~169142}
The fact that $^{13}$CO $J=6-5$ is detected just inside the dust cavity, but drops steeply inwards of $\sim$15~AU indicates that the gas cavity is likely very depleted in gas (Figs.~\ref{fig:aziavgmom0}, \ref{fig:kinfitter}). Therefore, we focus on the region outside $\sim10$~AU. 
The line ratio of $^{13}$CO $J=6-5$ to $^{13}$CO $J=2-1$ for HD~169142 increases to a value of $\sim$13 at the inner edge of the gas cavity and rapidly decreases further out in the disk (Fig.~\ref{fig:LkCa15ratio6521}) . 

The results of the line ratio analysis are shown in the right hand panels of Figs.~\ref{fig:Trot}, \ref{fig:tau}. The optical depths of the lines are comparable to the values found for LkCa15, i.e. the $^{13}$CO $J=2-1$ transition has an optical depth of a few. Furthermore, the temperature in the outer disks of HD~169142 and LkCa15 are similar at $\sim30$~K as derived from the line ratio. This derived temperature in the outer disk of HD~169142 is likely very low due to the small MRS of the Band~6 and Band~9 data, which lowers the $^{13}$CO $J=6-5$ and $J=2-1$ integrated intensities in the outer disk outside $\sim35$~AU. Furthermore, at the location of the bright continuum rings, the temperature from the line ratio may be affected by the continuum subtraction of the $^{13}$CO $J=6-5$ and $J=2-1$ lines used for the line ratio. The bright rim just inside the dust cavity of the HD~169142 disk is warm and (marginally) optically thick in $^{13}$CO $J=6-5$. The line ratio analysis predicts an increase to $\sim50$~K compared to the outer disk, but the uncertainty on the line ratio and optical depth are large. In contrast with the LkCa15 case, the $^{12}$CO $J=2-1$ brightness temperature suggests that the dust cavity is much warmer, up to 170~K, compared to the outer disk ($\sim100$~K).

The total hydrogen column density, derived using the rotational temperature and the integrated C$^{18}$O intensity (purple line in Fig.~\ref{fig:NHtot}, right panel), shows similar values at the location of the two rings detected in this disk and a few times higher than in the LkCa15 disk. The HD~169142 column density at 23~AU a factor of 2.5 lower than that modelled by \citet{Fedele2017}. This difference is due to the lower spatial resolution and larger MRS of the data used in that work. The drop in the gas column density by a factor 40 in the gap at 56~AU found by \citet{Fedele2017} is not constrained in our work due to the small MRS. 
The signal-to-noise ratio and MRS of the data are not high enough to determine the expected steady decrease in gas column density in the outer disk outside of $\sim35$~AU.

 \begin{figure*}
   \centering
  \begin{subfigure}{0.99\textwidth}
  \centering
  \includegraphics[width=1\linewidth]{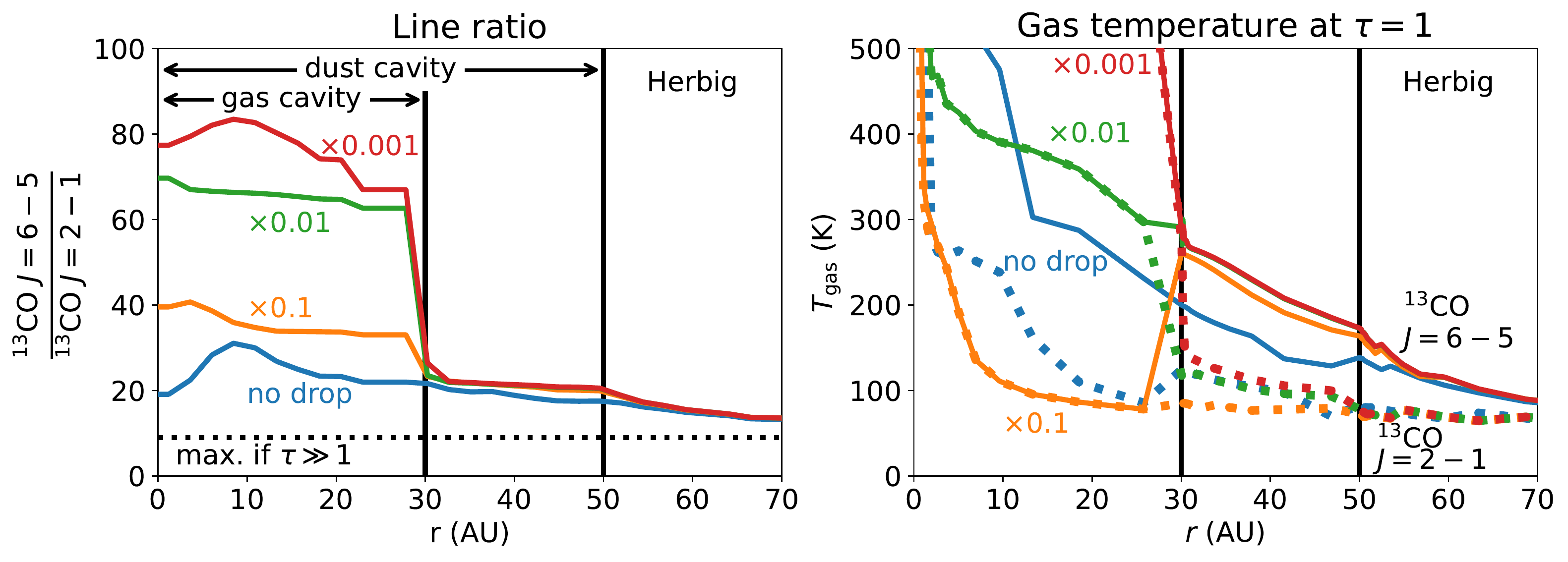}
\end{subfigure}%

\begin{subfigure}{0.99\textwidth}
  \centering
    \includegraphics[width=1\linewidth]{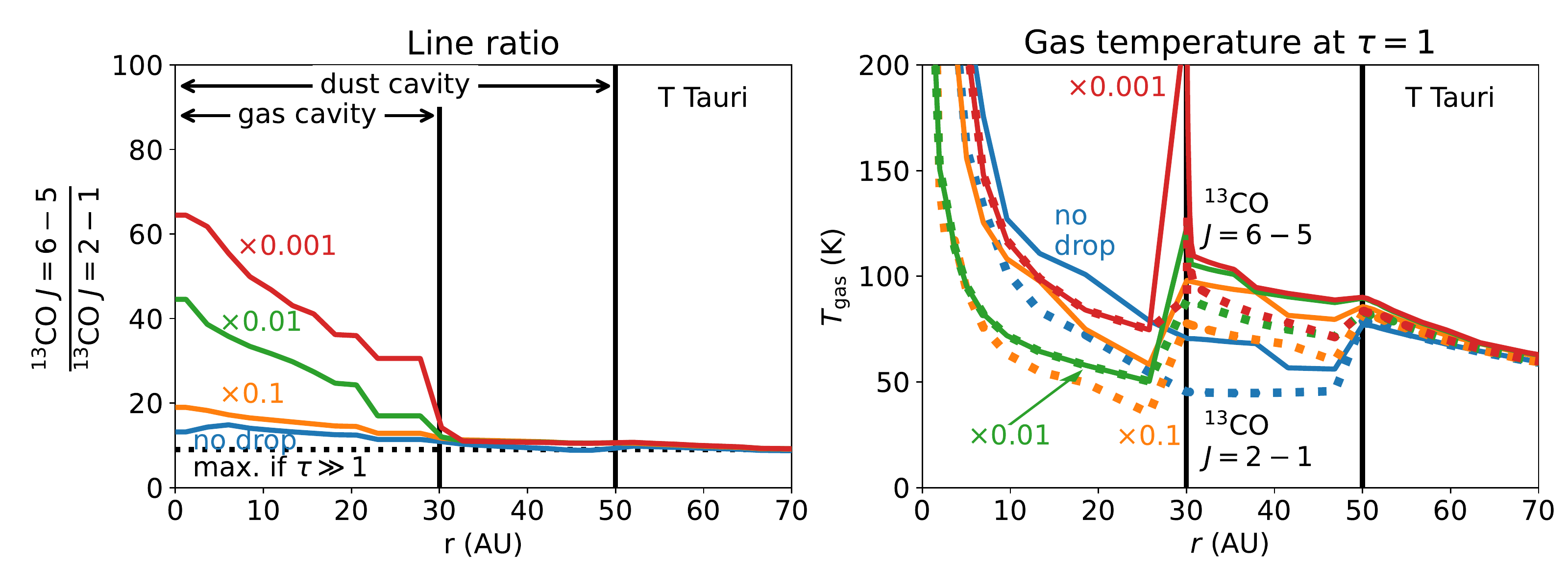}
\end{subfigure}
      \caption{Predicted line ratio (left) and gas temperature (right) at the height where the $^{13}$CO $J=6-5$ (solid) and $J=2-1$ (dotted) transitions become optically thick for different gas cavity depth (no drop to factor of 0.001, colors) in the Herbig disk models (top) and T~Tauri models (bottom). Note the different temperature scales for the two types of stars. If the transitions are optically thin, the midplane temperature is shown, as that is the region that optically thin CO lines trace in the gas cavity. The dotted horizontal line in the left panels indicates the theoretical maximum value for the line ratio if both $^{13}$CO lines are optically thick and emit from the same disk layer.  }
         \label{fig:lineratiodali}
   \end{figure*}

\section{Thermochemical models} \label{sec:dalisection}

The analysis in Sect.~\ref{sec:ana} has shown that the temperature increases inside the dust cavity compared to outside for both the LkCa15 and HD~169142 disks. Furthermore, it was found that the gas is located at smaller radii, $\sim10$~AU, than the dust cavity and that the gas cavity is very deep ($\geq 100 \times$ drop in gas column density) in the HD~169142 disk. The gas in the dust cavity for LkCa15 could be optically thin, but only an upper limit on the column density could be obtained due to the beam size being comparable to the dust cavity and because the line ratio exceeded the maximum possible value of 9 if both lines are optically thick. Finally, a difference between the temperature derived from the line ratio and the brightness temperature was found. Although this could be due to the MRS being smaller than the angular size of the gas and dust disk in the case of HD~169142, another explanation would be a difference in the emitting layers of the $^{13}$CO $J=6-5$ and $J=2-1$ transitions. This is in line with the expectation that different lines do not emit from the same disk region due to optical depth and excitation effects (e.g., \citealt{vanZadelhoff2001}) as also derived directly in the LkCa15 disk (see Fig.~\ref{fig:emittingheights}).
In order to investigate this situation, we present some exploratory DALI models following \citet{Bruderer2013}. Then we compare our results to other sources and finally discuss the implications.

 \begin{figure*}
   \centering
    \includegraphics[width=2\columnwidth]{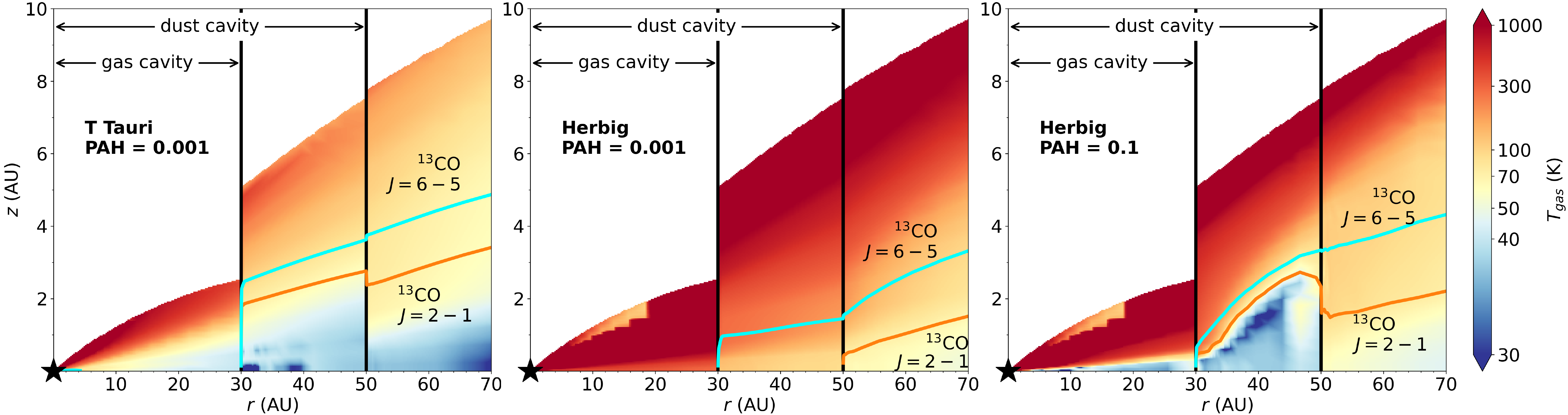}
      \caption{Gas temperature of our model for a T~Tauri star and a PAH abundance of 0.001 w.r.t. the ISM (left), the fiducial Herbig model with a PAH abundance of 0.001 w.r.t. the ISM, and a Herbig model with a high PAH abundance of $0.1$ w.r.t. the ISM (right). All models have a factor of 100 drop in gas density inside 30~AU. The cyan and orange contours indicate the $\tau = 1$ surface of the $^{13}$CO $J=6-5$ and $J=2-1$ lines respectively. The $^{13}$CO $J=2-1$ transition is optically thin inside the dust cavity and the $J=6-5$ is optically thin inside the gas cavity, hence these lines trace the midplane at these radii. Note that the dust cavity starts from 50~AU inwards and the gas cavity from 30~AU. Only the regions with a gas density above $10^7$~cm$^{-3}$ are shown.  
       }
         \label{fig:DALImodelLkCa15T}
   \end{figure*}

The thermochemical code DALI is used to model the line ratio of the $^{13}$CO $J=6-5$ to $J=2-1$ transition in the cavities of typical transition disks. A detailed description of the DALI modelling setup is given in Appendix~\ref{app:dali} and in Table~\ref{tab:paramsdali}. We do not attempt to model the individual sources, but look for trends in a set of representative disk models and study the sensitivity to parameters. 
The fiducial Herbig disk model is a transition disk with a gas cavity with variable gas drop inside 30~AU. The dust cavity extends to 50~AU and a small inner disk of 1~AU is present. The ALMA observations presented in this work mainly probe the region inside the dust cavity but outside the gas cavity corresponding to $30-50$~AU in this model. The density and temperature structure of this model are presented in Fig.~\ref{fig:DALImodelLkCa15n}. 
The cyan and orange lines show the $\tau = 1$ surfaces of the $^{13}$CO $J=6-5$ and $J=2-1$ transitions respectively. For gas density drops by factors of $0.1-0.001$, both $^{13}$CO lines are (marginally) optically thin in the gas cavity. Outside the gas cavity and in the model with no gas drop in the dust cavity, both lines are optically thick and trace a slightly different layer in the disk. 

The predicted $^{13}$CO $J=6-5$ to $J=2-1$ line ratios for this model with four different gas drops inside 30~AU are presented in the top left hand panel of Fig.~\ref{fig:lineratiodali}. This figure shows that the line ratio can reach up to the maximum possible value of 80 for optically thin gas in the gas cavity, due to the high gas temperatures in the gas cavity midplane and both $^{13}$CO lines being optically thin. The temperature in the gas cavity midplane increases due to the increased UV radiation field if less gas is present. Note that PAHs are present in the gas even if at small amounts and absorb the UV radiation from the star. This increased radiation field for lower gas densities also changes the chemistry in the cavity by photodissociating CO, the major coolant at a height of $\sim1$~AU inside the gas cavity. Therefore, the gas inside 30~AU in the models with a deep gas cavity, cools less efficiently than in the full gas disk model. This results in an increased midplane temperature inside 30~AU with deeper gas cavity depth (compare $0.1-0.001\times$ drops in Fig.~\ref{fig:lineratiodali} right hand panel). In these models with a gas density drop ($\geq 10\times$), the line ratio is a direct probe of the temperature. 

Outside the gas cavity but inside the dust cavity and in the outer disk ($R>30$~AU), the line ratio is larger than 9 which is the maximum theoretical value that is expected when the emission is optically thick. This is because the $J=6-5$ line emits from a layer with a higher temperature than the $J=2-1$ line (bottom panel Fig.~\ref{fig:DALImodelLkCa15n} and \ref{fig:lineratiodali}, top right panel), increasing the line ratio w.r.t. the $J=2-1$ transition up to 20 in the outer disk.
In summary, the line ratio analysis can be used if both lines are emitting from the same disk region.
If the lines become optically thick, their emitting regions will deviate, and the brightness temperature becomes a good probe of the temperature of that layer.

The effect of the host star is investigated as LkCa15 is a 1.1~L$_{\odot}$ T~Tauri star and HD~169142 is a 10~L$_{\odot}$ Herbig star. The higher luminosity of the Herbig star heats the surrounding disk to higher temperatures as shown in the left two panels of Fig.~\ref{fig:DALImodelLkCa15T}. One of the main heating agents in the disk is molecular hydrogen that is vibrationally excited by the absorption of an FUV photon. The collisional de-excitation then releases this energy to the gas causing it to heat up. This process is most effective in the gas cavity midplane, and at $z/r\sim 0.1$ for radii $>$30~AU. The disk midplane in the T~Tauri model is colder than the corresponding region in the Herbig model because the main heating agents, FUV pumped H$_2$ and the photoelectric effect on small grains and PAH, are less effective due to the weaker FUV field and lower luminosity in the T~Tauri model.

The line ratio and temperature at the $\tau=1$ surface for the T~Tauri models are presented in the bottom row of Fig.~\ref{fig:lineratiodali}. The line ratios in the T~Tauri disk models are smaller than the values in the Herbig disk models. This is due to the lower gas temperature in the T~Tauri models (Fig.~\ref{fig:lineratiodali}, right hand panel). Another difference between the Herbig and the T~Tauri models is the temperature of the emitting layers. In the outer disk of the Herbig models, the temperature of the $^{13}$CO $J=6-5$ layer is $\sim1.3-2\times$ higher than in the $J=2-1$ layer, but in the T~Tauri models the layers have nearly the same temperature of $50-100$~K. Inside the dust cavity, the temperatures of the $J=6-5$ and $J=2-1$ layers differ from each other except within the 30~AU gas cavity for the models with a deep gas cavity (drop $\geq 0.01$), as the lines become optically thin in this region. The effects of the PAH abundance, the presence of a dusty inner disk and the effects of grain growth are investigated in Appendix~\ref{sec:daliother}.

The exploratory DALI models show that the UV field and the amount of gas in the gas cavity are important for the temperature of the gas in the cavity. The disk models by \citet{Bruderer2009, Bruderer2012, Bruderer2013, Alarcon2020} do not distinguish between a gas and dust cavity or gap and show that the gas in this region is warmer than the gas outside the cavity or gap. In contrast, the models by \citet{Facchini2017, Facchini2018} show lower temperatures. The main difference between these models are the heating and cooling rates that are taken to be independent of grain sizes and gas-to-dust ratios in the cavities/ gaps in the models by \citet{Bruderer2013}, whereas they are dependent on grain sizes in the models by \citet{Facchini2017} and in this work. In our models, we show that the gas temperature not only depends on these choices, but also on the stellar luminosity (Fig.~\ref{fig:DALImodelLkCa15T}). In our T~Tauri models with a gas cavity the emitting layer of $^{13}$CO $J=6-5$ and $J=2-1$ inside the dust cavities ($10-50$~AU) have a temperature similar to or colder than the outer disk, whereas in the Herbig disk models, we find that the temperature of the emitting layers is generally warmer than the outer disk.

\section{Discussion} \label{sec:disc2}
\subsection{Comparing models and observations}

The DALI models show that the $^{13}$CO $J=6-5$ to $J=2-1$ line ratio is a sensitive tracer of the gas temperature in the gas cavity of transition disks if the emission is optically thin. Outside the gas cavity and in the outer disk regions of the models, the two $^{13}$CO lines are optically thick, complicating the line ratio analysis, but in this case the brightness temperature can be used to constrain the temperature. The line ratio inside the gas cavity is especially sensitive to the amount of gas in the gas cavity, the presence of a dusty inner disk, the abundance of small grains and PAHs in the gas cavity, as these are all important parameters to set the temperature. A higher line ratio points towards a low PAH abundance, no dusty inner disk, and grain growth. 

Our ALMA observations in the HD~169142 disk trace the disk region outside the gas cavity but inside the dust cavity (from 30 to 50~AU in the models and from $5-10$ to 25~AU in the data). In this region, the line ratio is not very sensitive to the amount of gas in the gas cavity due to optical depth effects and because its temperature structure is not affected much when a gas cavity is introduced. In contrast, the line ratio in this disk region is affected by the presence of a dusty inner disk, the abundance of PAHs and grain growth similar to the disk region inside the gas cavity. Therefore, the ALMA line ratio in the HD~169142 disk mainly trace the effects of these parameters rather than the amount of gas in the gas cavity. 

The $^{12}$CO brightness temperature indicates that the gas is 170~K in this region in the HD~169142 disk, which is consistent with the $80-300$~K temperature range predicted by the DALI models inside the dust cavity but outside the gas cavity. Similarly, the observed brightness temperature of $\sim$100~K in the outer disk of HD~169142 is consistent with the typical temperature of the $^{13}$CO emitting layers in the fiducial Herbig disk model. 

In the case of LkCa15, the gas cavity of $\sim10$~AU is unresolved by ALMA observations. Therefore, the  observed line ratio likely traces a combination of all effects (gas cavity depth, abundance of PAHs, presence of a dusty inner disk and grain growth) discussed above, complicating the interpretation of the observations. The T~Tauri models have a typical temperature of $50-100$~K in the outer disk at their $\tau=1$ heights, which is higher than the typical temperature of $20-30$~K in the outer disk of LkCa15. Still, the models predict a steep increase in the gas temperature inside the gas cavity. This is likely also traced by our observations, but the large beam smears this out to larger radii up to the dust cavity radius.

\subsection{Comparison to other sources}

The temperature structure of few other sources has been studied in detail. Recent work on the temperature structure in the transition disk GM~Aur has shown that an additional increase of a factor of up to 10 in the gas temperature in the surface layers in the inner disk, inside its dust cavity at 40~AU, compared to the thermochemical model prediction is needed to fit the observed intensities \citep{Schwarz2021}. In contrast, that same model overpredicts the gas temperature in the outer disk by a factor of 2. The increase in temperature inside the GM~Aur dust cavity is similar to the steep increase in gas temperature inside the dust cavity found in the LkCa15 and HD~169142 disks. One of their proposed explanations is that the PAH abundance in the surface layer is too low in the GM~Aur models. Our models confirm that the abundance of PAHs is important together with the UV field to effectively heat the gas through the photoelectric effect. On the other hand, a high PAH abundance also shields the gas from stellar UV radiation causing a lower gas temperature in the emitting layer. The disk region inside the inner most beam for LkCa15 may even be warmer than what we derived using the $^{13}$CO $J=6-5$ to $J=2-1$ line ratio, because of beam dilution. This effect is less relevant for HD~169142 as its gas cavity is very depleted in gas. 

{High resolution $^{13}$CO $J=6-5$ observations probing the inner few AU in the TW~Hya disk are not available. The temperature in the outer disk of TW Hya, derived using a similar method to this work, it is also consistent with excitation temperature derived from formaldehyde and methyl cyanide emission \citep{Schwarz2016, Loomis2018, Pegues2020, Jeroen2021}. As all of these temperatures are similar to each other, the molecules used likely emit from similar disk regions as the disk is very flat. This situation is similar to our case for LkCa15 which has similar outer disk temperatures as the TW Hya disk and also a shallow vertical temperature gradient.

Modelling by \citet{Fedele2016} has shown that the temperature in four Herbig disks depends on their vertical structure (flaring and scale height) as well as the dust distribution (gas-to-dust ratio and dust settling). Comparing the temperature of the HD~169142 disk to the temperatures found in Herbigs using  \textit{Herschel} observations of the CO ladder shows that HD~169142 is $\sim50-100$~K warmer than the HD~163296 disk at the emitting heights. Furthermore, HD~169142 falls within the range of all four Herbig disks analysed in \citet{Fedele2016}.

\subsection{Implications for the cause of the gas cavity}

Our results have shown that temperature plays a role in lowering the low-$J$ lines. One of the most popular alternative explanations for cavities and rings in protoplanetary disks are planets. The deep gas cavity in the HD~169142 disk could be carved by a massive companion. The high signal-to-noise ratio of the $^{12}$CO $J=2-1$ transition constrains the total hydrogen column density in the inner 10~AU to be $\leq (3-7)\times 10^{20}$~cm$^{-2}$ using a $3\sigma$ upper limit on the integrated $^{12}$CO $J=2-1$ intensity and assuming an average gas temperature of 100$-$300~K, respectively. This upper limit shows that the deep gas cavity that is depleted by at least a factor of $(2-5)\times 10^2$, w.r.t. the bright ring seen in the gas. If this cavity is carved by a massive companion, its mass can be estimated using the following equation 
\citep{Zhang2018DSHARP}:
\begin{align}
\frac{K}{24} = \frac{q}{0.001} \left (\frac{h/r}{0.07} \right )^{-2.81} \left ( \frac{\alpha}{10^{-3}}\right )^{-0.38},
\end{align}
with $q$ the mass ratio of the companion to the star, $\alpha$ the viscosity and $h/r$ the scale height of the disk. $K$ is a dimensionless parameter describing the depth of the gas cavity, $\delta$, as $\delta-1 = CK^D$. The values of $C = 0.002$ and $D=2.64$ are the best fitting parameters to the hydro simulations of both the gas and the dust performed by \citet{Zhang2018DSHARP}. 
\citet{Fedele2017} found from models that the disk scale height $h/r$ is 0.07 independent of radius. Assuming a typical disk viscosity of $10^{-3}$ constrains the mass of the potential companion to be $\geq8.0$~M$_{\mathrm{J}}$ for a gas temperature of 100~K in the gas cavity and $\geq5.5$~M$_{\mathrm{J}}$ for 300~K. The mass of the potential planet and the disk viscosity are degenerate. Therefore, a lower disk viscosity of $10^{-4}$, results in the mass of the potential planet to be $\geq 3.3$~M$_{\mathrm{J}}$ (100~K) or $\geq 2.3$~M$_{\mathrm{J}}$ (300~K). A cartoon of the proposed scenario is presented in Fig.~\ref{fig:cartoon}. It is also possible that multiple smaller planets carve the cavity seen in the gas \citep{DodsonRobinson2011}.

LkCa15 may be hosting multiple low mass planets, as the gas cavity ($\sim10$~AU) is much smaller than the dust cavity (68~AU). 
In LkCa15, the gas column density drops by at least a factor of 2 inside the dust cavity, though the exact  drop cannot be inferred from the observations due to large beam of the C$^{18}$O data. Still, the CO column density inside 0.3~AU derived by \citet{Salyk2009} constrains the total drop in gas column density inside the gas cavity w.r.t. the outer disk to be a factor 24. The gas detection up to $\sim10$~AU points to the presence of multiple low mass planets that are massive enough to carve a cavity in the dust up to 68~AU but not (yet) in the gas \citep{Dong2015}, see Fig.~\ref{fig:cartoon}.

Another possible mechanism to create cavities in transition disks is internal photoevaporation. In this case low accretion rates of $\lesssim 10^{-9}$~M$_{\odot}$~yr$^{-1}$ are expected, as the gas is evaporated from the cavity, preventing it to be accreted onto the central star \citep{Owen2011, Picogna2019, Ercolano2021}. 
The mass-accretion rate of $10^{-9.2}~\dot{M}$~M$_{\odot}$~yr$^{-1}$ for LkCa15 can be consistent with internal photoevaporation, but we find that the gas cavity of $\sim10$~AU is much smaller than the dust cavity of 68~AU, which is inconsistent with what is expected for a cavity carved by photoevaporation \citep{Donati2019}. These models predict that small grains are removed from the cavity together with the gas, while the larger grains remain present \citep{Franz2020}, such that the cavity seen in gas and small dust is larger than the cavity seen in large dust grains. This is the opposite of what is observed in the LkCa15 disk as the mm-sized grains peak at larger radii (68~AU) than the $\mu$m-sized grains (56~AU; \citealt{Thalmann2014}) and our gas data (10~AU).
In the case of the HD~169142 disk, the mass accretion rate of $10^{-7.4}~\mathrm{M_{\odot}\ yr^{-1}}$ \citep{Guzman-Diaz2021} is at least one order of magnitude larger than what can be explained by photoevaporation models.

Dead zones can also cause a drop in the gas column density in the inner disk, but the drops due to dead zones are expected to be on the order of a factor of a few \citep{Lyra2015}.
The deep drops in the gas column density inside the dust cavities of LkCa15 and HD~169142, together with the presence of gas well inside the dust cavity points to the presence of one or more lower mass planets that are hard to detect. Moreover, multiple planets can produce a gap that has a similar width to a gap with a massive planet, making their detection even more challenging.

  \begin{figure}
        \centering
        \includegraphics[width=\linewidth]{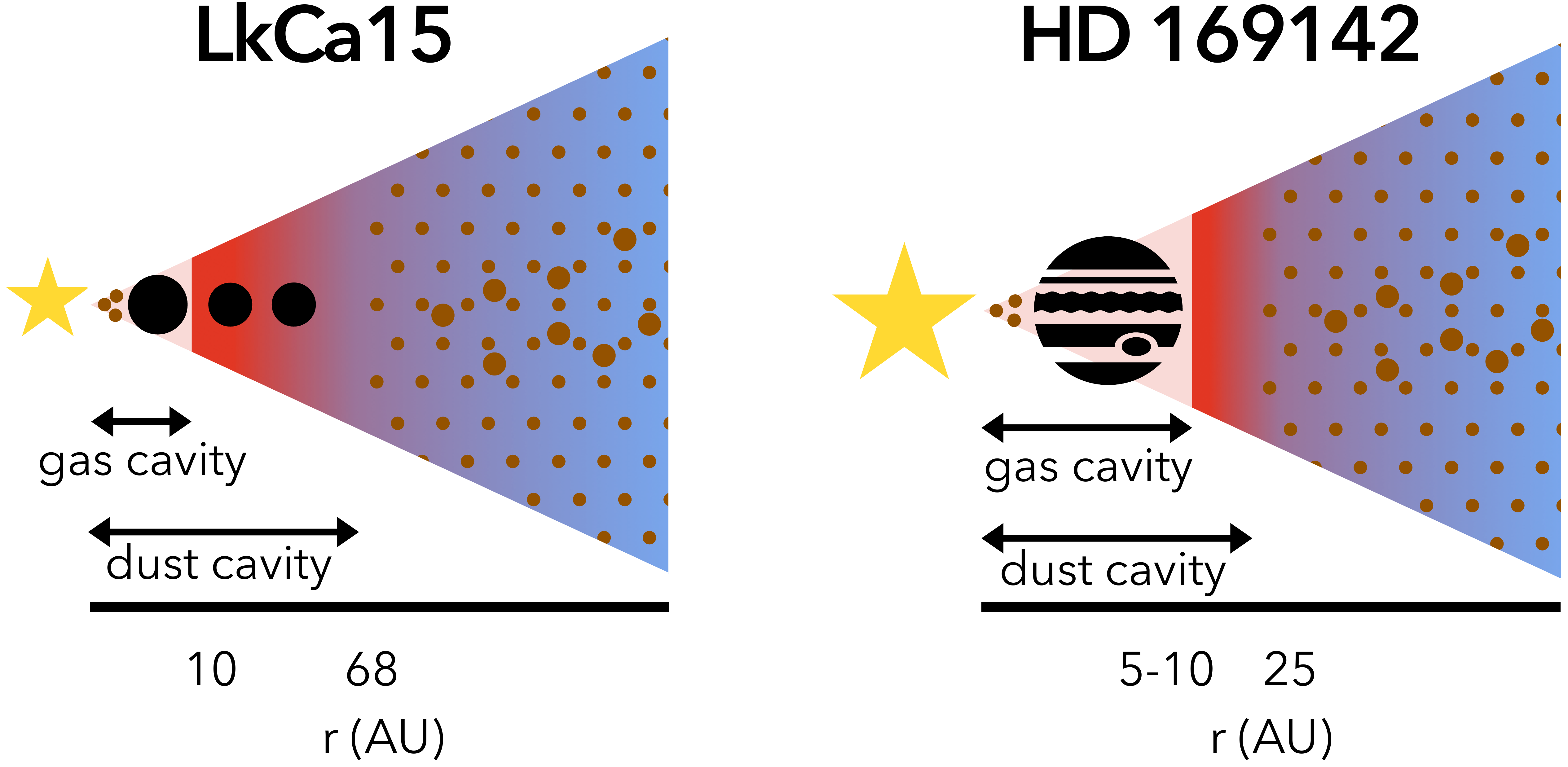}
        \caption{Cartoon of the proposed scenario for LkCa15 (left) and HD~169142 (right). The black circles represent the proposed planets that may be carving the dust cavities of both disks. The colored background indicates the radial gas temperature structure in both disks. The brown circles indicate where the dust is present. }
        \label{fig:cartoon}
\end{figure}

\section{Conclusions} \label{sec:concl}
In this work we used the observed $^{13}$CO $J=6-5$ to $J=2-1$ line ratio to determine the temperature in the cavities for LkCa15 and HD~169142 with ALMA. The line ratio in the LkCa15 and HD~169142 disks steeply increases inside the dust cavity towards the position of the star in the (combined) high resolution data. Outside the dust cavity, both disks have a relatively low ratio of $\lesssim 6$. 
The temperature is derived from the line ratio and brightness temperature analysis. Also DALI models are used to investigate the effect of a drop in the dust and gas column density in the dust and gas cavity on the line ratio. Furthermore, we also investigated the effect of the stellar spectrum, PAH abundance, grain growth and a dusty inner disk on the line ratio as these parameters affect the local UV field. Our conclusions are summarized below: 
\begin{enumerate}
      \item The gas cavity is smaller than dust cavity for both disks. Based on the kinematically derived radial profiles we find that the gas in the LkCa15 disk extends down to at least 10~AU, much further than the dust ring at 68~AU. The results for HD~169142 are consistent with a $\sim5-10$~AU gas cavity and a dust ring at 25~AU. 
      \item The integrated intensity of the $^{13}$CO $J=6-5$ transition is relatively flat in the dust cavity of the LkCa15 disk whereas the $J=2-1$ transition shows a steep decrease. Therefore, gas is present inside the dust cavity and the excitation is important to determine its column density. The molecular excitation can be constrained directly by the $^{13}$CO $J=6-5$ to $J=2-1$ line ratio if both lines are optically thin. 
      \item The line ratio of $^{13}$CO $J=6-5$ to $J=2-1$ reaches a high value of $13$ at the inner rim of the HD~169142 gas (15~AU) disk and it steeply increases towards the position of the star inside the dust cavity of the LkCa15 disk. This indicates that these disk regions are warm compared to the dusty outer disk. The dust cavity of the LkCa15 disk (up to $65$~K) is warmer than the outer disk ($\sim20-30$~K), based on the line ratio analysis. For HD~169142 this analysis is not possible as the lines are optically thick. 
      \item The brightness temperature of optically thick lines is a good alternative probe of the temperature across the cavity for HD~169142 and indicates that the gas in the dust cavity, up to 170~K, is warmer than in the outer disk, up to $\sim100$~K. For LkCa15, the brightness temperature is only a good probe of the temperature in the outer disk due to beam dilution.
      \item The factor of $(2-5)\times 10^2$ drop in gas column density in the HD~169142 cavity indicates that a massive ($\geq 2.3-8.0$~M$_{\mathrm{J}}$) planet may be present in this region. For LkCa15, the extended gas-rich region inside the dust cavity may be indicative of several lower mass planets.  
      \item The DALI models show that the $^{13}$CO $J=6-5$ and $J=2-1$ transition emit from different layers in the disk if the lines are optically thick. Their line ratio in the cavities of transition disks can reach high values up to the theoretical maximum value of 80 depending on the model. In general, models with a high luminosity, a low PAH abundance, no dusty inner disk and a large fraction of large grains predict the highest temperatures and line ratios. 
      \item The vertical structure inferred from the channel maps in the LkCa15 disk is very flat with a typical scale height of $z/r\sim0.1$ for the $^{13}$CO $J=2-1$ transition. Furthermore, the $^{13}$CO $J=6-5$ transition emits from a higher disk layer ($z/r\sim0.2-0.3$) than the $^{13}$CO $J=2-1$ transition, consistent with thermochemical models. 
\end{enumerate}

In summary, the brightness temperature is a good probe of the disk temperature if the (low-$J$) CO lines are optically thick (e.g. for HD~169142). If the lines are optically thin, the $^{13}$CO $J=6-5$ to $J=2-1$ line ratio is a sensitive probe of the gas temperature. The intermediate case, where these lines are marginally optically thin or not very optically thick, the analysis is more difficult (e.g. for LkCa15). The line ratio analysis in LkCa15 clearly shows that an intensity drop in a low-$J$ CO line does not necessarily require a drop in the CO column density. This work also highlights that resolving the cavities with more than one beam and with good sampling of the relevant scales is crucial for the analysis. High spatial resolution ALMA observations of high-$J$ CO isotopologs (both optically thick and thin) of a much larger sample of transition disks is needed to nail down the combined temperature and column density structure of the gas inside dust cavities, and thus, ultimately, the mechanism responsible for carving them.

\begin{acknowledgements}
We thank the referee for the constructive comments. Astrochemistry in Leiden is supported by the Netherlands Research School for Astronomy (NOVA), by funding from the European Research Council (ERC) under the European Union’s Horizon 2020 research and innovation programme (grant agreement No. 101019751 MOLDISK), and by the Dutch Research Council (NWO) grants 618.000.001 and TOP-1 614.001.751. Support by the Danish National Research Foundation through the Center
of Excellence “InterCat” (Grant agreement no.: DNRF150) is also acknowledged. A.D.B. acknowledges support from NSF AAG grant No. 1907653. This paper makes use of the following ALMA data: ADS/JAO.ALMA \#2016.1.00344.S, \#2017.1.00727.S, \#2018.1.00945.S, and \#2018.1.01255.S. ALMA is a partnership of ESO (representing its member states), NSF (USA) and NINS (Japan), together with NRC (Canada), MOST and ASIAA (Taiwan), and KASI (Republic of Korea), in cooperation with the Republic of Chile. The Joint ALMA Observatory is operated by ESO, AUI/NRAO and NAOJ. 
\end{acknowledgements}

\bibliographystyle{aa}
\bibliography{refs.bib}

\begin{appendix}

\section{Observations} \label{app:obs}
\subsection{Channel maps $^{13}$CO $J=6-5$} \label{app:chanmaps}

The channel maps of the $^{13}$CO $J=6-5$ transition in the LkCa15 and HD~169142 disks are presented in Fig.~\ref{fig:chanmaps} and \ref{fig:chanmapsHD}, respectively. 

    \begin{figure*}
        \centering
        \includegraphics[width=\textwidth]{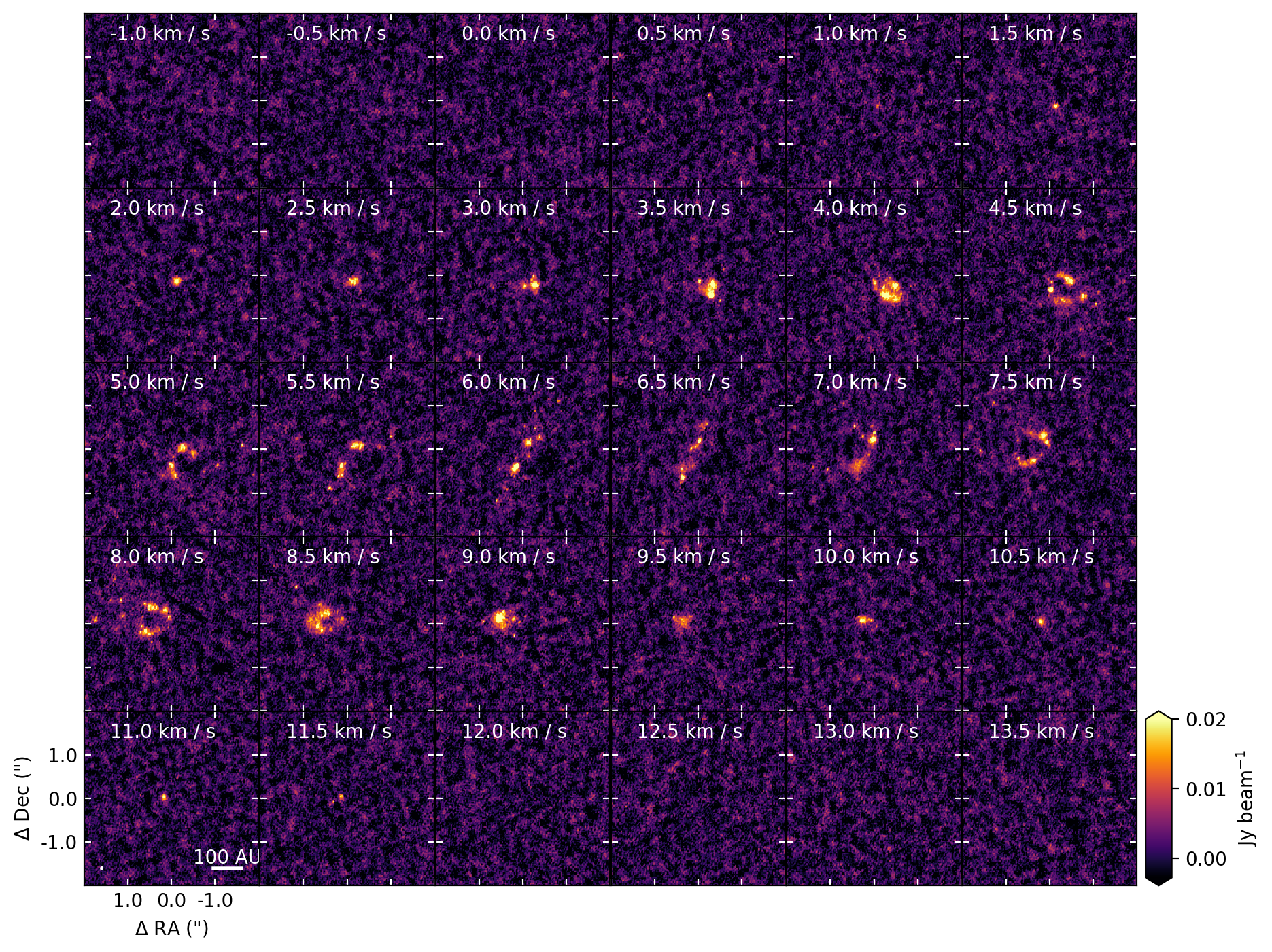}
        \caption{Channel maps of the combined dataset for the $^{13}$CO $J=6-5$ transition in the LkCa15 disk.}
        \label{fig:chanmaps}
    \end{figure*}

    \begin{figure*}
        \centering
        \includegraphics[width=\textwidth]{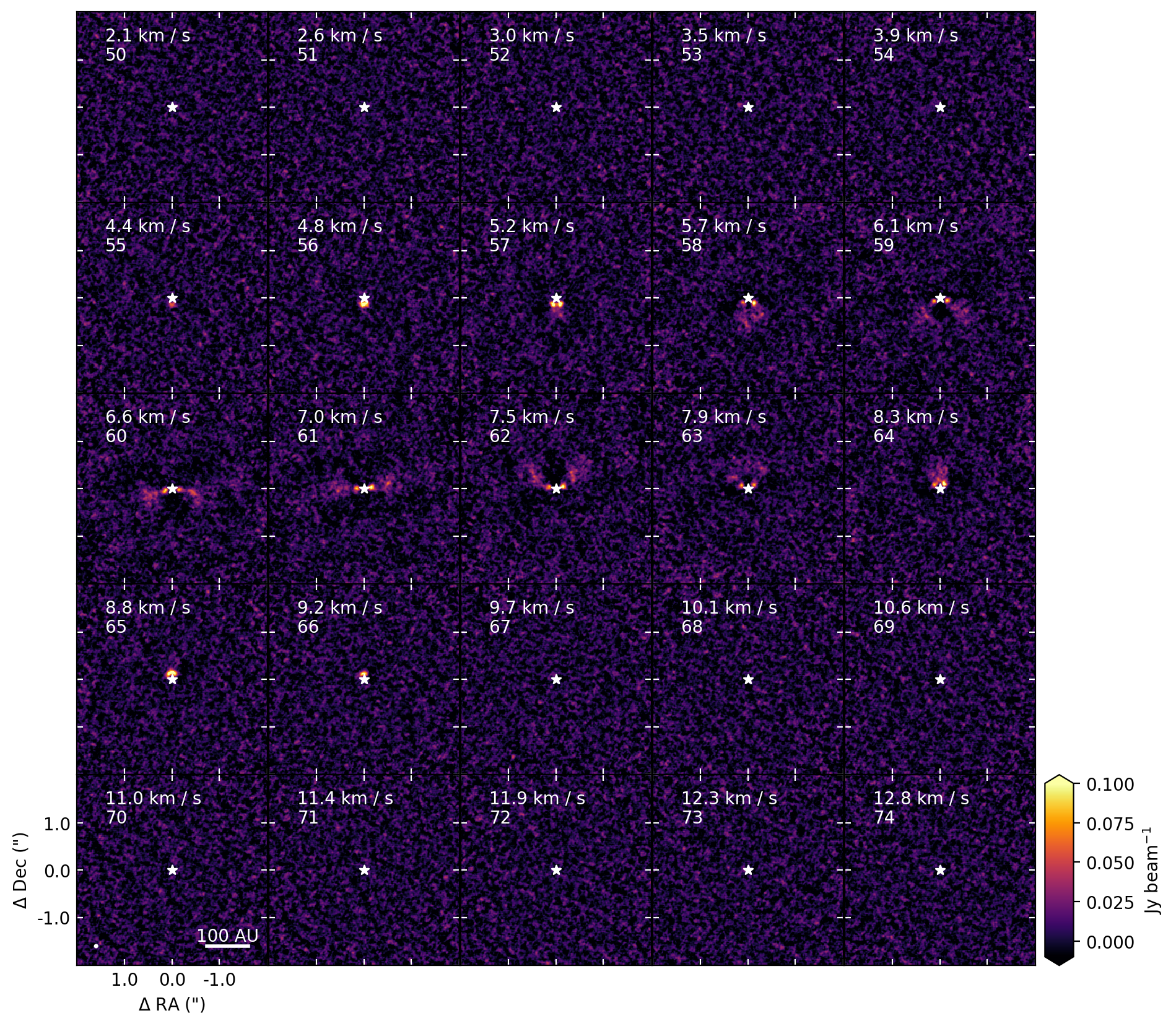}
        \caption{Same as Fig.~\ref{fig:chanmaps} but then in the HD~169142 disk.}
        \label{fig:chanmapsHD}
    \end{figure*}

\subsection{Azimuthally averaged radial profiles} \label{app:aziavg}

The deprojected azimuthal averages of the CO isotopologue emission are presented in Fig.~\ref{fig:aziavgmom0app}.

    \begin{figure*}
        \centering
        \includegraphics[width=\textwidth]{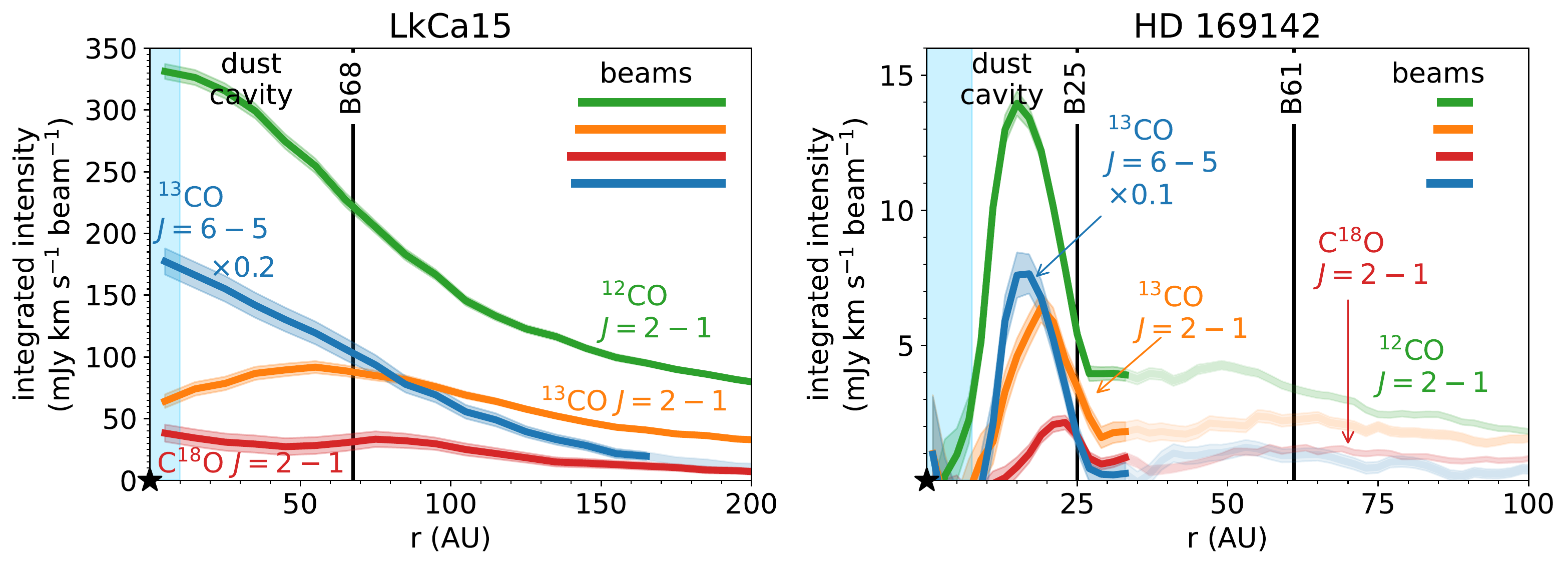}
        \caption{Azimuthally averaged radial profiles of $^{12}$CO, $^{13}$CO, and C$^{18}$O $J=2-1$ and $^{13}$CO $J=6-5$ at intermediate resolution for LkCa15 (left) and at high resolution for HD~169142 (right). The gas cavity (light blue shaded region) is unresolved at these spatial resolutions. Left: we note that the morphology of the $^{13}$CO $J=6-5$ is very different from the $^{13}$CO $J=2-1$ line. Furthermore, the integrated $^{13}$CO $J=6-5$ line intensity was multiplied by $0.2$ (LkCa15) and $0.1$ (HD~169142) for visibility. The (combined) high resolution data for the LkCa15 disk is presented in Fig.~\ref{fig:aziavgmom0}.}
        \label{fig:aziavgmom0app}
    \end{figure*}

\subsection{Brightness temperature} \label{app:Tbspec}  
The brightness temperature of an optically thick line is calculated using the peak flux. However, a low spectral resolution can lower this peak flux compared to the true value. In this Section, we quantify this effect for the $^{13}$CO emission. 
The spectral resolution of the dataset covering the $^{13}$CO $J=2-1$ line is 0.17~km~s$^{-1}$, whereas that of the $J=6-5$ dataset is 0.44~km~s$^{-1}$ in HD~169142. Therefore, the former dataset better samples the line profile and hence the peak flux of the $^{13}$CO emission than the latter. The effect of underestimating the peak flux due to a lower spectral resolution is shown in Fig.~\ref{fig:Tbspec}, where the brightness temperature of the $^{13}$CO $J=2-1$ transition is shown at two spectral resolutions: 0.17~km~s$^{-1}$ (native resolution solid orange line) and 0.44~km~s$^{-1}$ (binned, dashed orange line). This figure shows that binning lowers the brightness temperature by 10-15~K, corresponding to $\sim$10-30\%, compared to the brightness temperature at the native spectral resolution. For comparison the brightness temperature of the $^{13}$CO $J=6-5$, at 0.44~km~s$^{-1}$ is shown as the blue dashed line. This figure shows that the low spectral resolution of the $^{13}$CO $J=6-5$ data is partially responsible for the difference between the $J=2-1$ and $J=6-5$ brightness temperatures discussed in Section~\ref{sec:Tb}. 
Similarly, in the LkCa15 disk, the lowering the spectral resolution of the high spatial resolution $^{13}$CO $J=2-1$ line from 0.17~km~s$^{-1}$ to 0.5~km~s$^{-1}$, lowers the peak brightness temperature by 6-13~K or 40-60\%. Therefore, the $^{13}$CO $J=6-5$ brightness temperature in LkCa15 is likely underestimated due to the spectral resolution.

\begin{figure*}
        \centering
        \includegraphics[width=\textwidth]{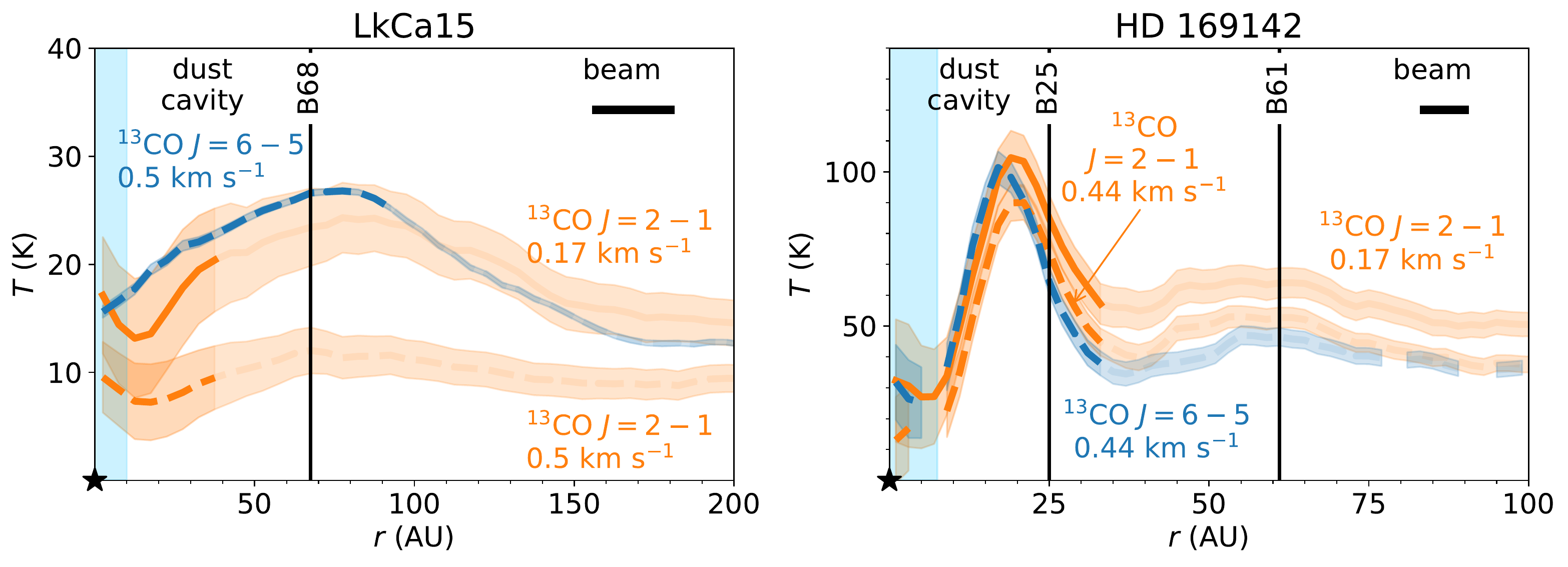}
        \caption{Deprojected azimuthally averaged radial profiles of brightness temperature of the $^{13}$CO $J=2-1$ (orange), and $^{13}$CO $J=6-5$ (blue) transition in LkCa15 (left) and  HD~169142 (right). The dashed lines show these at a spectral resolution of 0.5~km~s$^{-1}$ (LkCa15) and 0.44~km~s$^{-1}$ (HD~169142). The solid line shows the $^{13}$CO $J=2-1$ transition at the native spectral resolution of 0.17~km~s$^{-1}$. The difference between the solid and dashed orange lines is due to the spectral resolution as the cubes are convolved to a beam of $\sim0\farcs17\times0\farcs13\ (-13.7\degree)$ (LkCa15) and $0\farcs057\times0\farcs054\ (88.6\degree)$ (HD~169142), respectively. The vertical black lines indicate the rings seen in the Band~9 continuum and the light blue region starting from 0~AU indicates the gas cavity. The MRS of the HD~169142 $J=6-5$ data corresponds to a radius of 35~AU.}
        \label{fig:Tbspec}
\end{figure*}

\subsection{Spectra}

The signal-to-noise ratio of the spectra used for the kinematically derived radial profiles is presented in Fig.~\ref{fig:kinfitter}. The signal-to-noise ratio is the highest for the $^{12}$CO $J=2-1$ and $^{13}$CO $J=6-5$ transitions in the line wings, therefore these lines are also detected out to the smallest radii.

\begin{figure*}
        \centering
             \includegraphics[width=\textwidth]{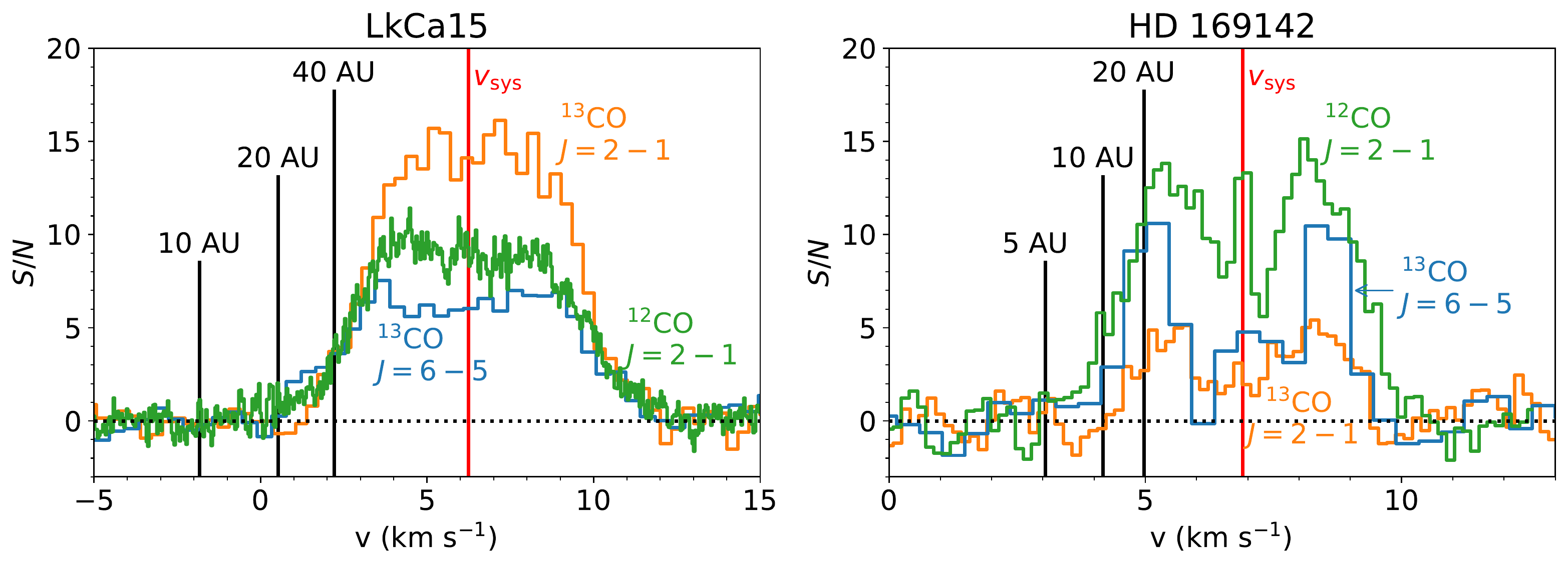}
        \caption{Signal-to-noise ratio of the $^{12}$CO $J=2-1$, $^{13}$CO $J=2-1$, and $^{13}$CO $J=6-5$ spectra in the inner $1\farcs0$ and $0\farcs5$ for LkCa15 (left) and HD~169142 (right), resp. The systematic velocity is indicated with the red vertical line. The vertical black line indicate the maximal velocities where emission at that radius is expected based on the Keplerian velocity profile. Note the difference in horizontal axes.}
        \label{fig:kinfitterspec}
   
\end{figure*}

\section{Line ratio analysis} \label{app:rotdia}
\subsection{Temperature and column density}

Rotational diagrams can be used to derive the temperature and column density of the gas observationally (e.g. \citealt{Goldsmith1999, Schwarz2016, Loomis2018, Pegues2020, Jeroen2021}). The relation between the column density and rotational temperature, as discussed in Section~\ref{sec:ana}, can best be solved in $\log$-space:
\begin{align}
\ln \left (\frac{N_u}{g_u} \right )  &= \ln N_{\mathrm{tot}} - \ln Q(T_{\mathrm{rot}}) -\frac{E_u}{kT_{\mathrm{rot}}} \\
\ln \left (\frac{N_u^{\mathrm{thin}}}{g_u} \right ) + \ln C_{\tau}  &= \ln N_{\mathrm{tot}} - \ln Q(T_{\mathrm{rot}}) -\frac{E_u}{kT_{\mathrm{rot}}}, \label{eq:rotdia}
\end{align}
where we used the correction factor for the optical depth in the last line. The optical depth is discussed in the next Section. 

\subsection{Optical depth} \label{app:tau}

The observations cover both $^{13}$CO $J=2-1$ and C$^{18}$O $J=2-1$. Therefore, the ratio of these two lines can be used to derive the optical depth of $^{13}$CO $J=2-1$. The equation of radiative transfer can be written as:
\begin{align}
T_{\mathrm{b}} &= T_{\mathrm{ex}}(1-e^{-\tau_{\nu}}),
\end{align}
with $T_{\mathrm{b}}$ the brightness temperature obtained from the non-continuum subtracted map and $T_{\mathrm{ex}}$ the excitation temperature of the line. The excitation temperatures of $^{13}$CO $J=2-1$ and C$^{18}$O $J=2-1$ are similar if both emit from the same region of the disk. Therefore, we can write:
\begin{align}
\frac{T_{\mathrm{b, 13CO\ 2-1}}}{T_{\mathrm{b, C18O\ 2-1}}} &= \frac{1-e^{-\tau_{\mathrm{13CO\ 2-1}}}}{1-e^{-\tau_{\mathrm{C18O\ 2-1}}}}.
\end{align}
As the line widths of the $^{13}$CO and C$^{18}$O $J=2-1$ transitions are similar, this line ratio is similar to the ratio of their integrated intensities, $I_{\nu}\Delta v$: 
\begin{align}
\frac{I_{\mathrm{13CO\ 2-1}}\Delta v}{I_{\mathrm{C18O\ 2-1}}\Delta v} &= \frac{1-e^{-\tau_{\mathrm{13CO\ 2-1}}}}{1-e^{-\tau_{\mathrm{C18O\ 2-1}}}}. \label{eq:tau} 
\end{align}
Furthermore, the ratio of the optical depths of $^{13}$CO and C$^{18}$O is identical to the isotopologue ratio, which we assume to be 8 \citep{Wilson1999}. The isotopologue ratio for $^{12}$CO to $^{13}$CO is assumed to be 70 \citep{Milam2005}. The uncertainty on the optical depths is dominated by the uncertainty on the line ratio of the $^{13}$CO to the C$^{18}$O $J=2-1$ transition. Assuming a 10~\% higher $^{12}$C/$^{13}$C ratio of 77, lowers the optical depths of the $^{13}$CO and C$^{18}$O $J=2-1$ transitions slightly but the values remain consistent within the error bars in the disk region where C$^{18}$O is detected inside the MRS. Therefore, the uncertainty on the isotopologue ratio is neglected.

The optical depth of the $^{13}$CO $J=6-5$ line can not be estimated in this way as no C$^{18}$O $J=6-5$ observations are available. Therefore, we use an analytical expression for the optical depth \citep{Goldsmith1999}:
\begin{align}
\tau_{\nu} &= \frac{A_{ul}c^3}{8\pi\nu^3\Delta V} \left (e^{h\nu/kT}-1\right )N_u.
\end{align}
This can be related to the optical depth of $^{13}$CO $J=2-1$, by taking the ratio of the two optical depths:
\begin{align}
\frac{\tau_{\mathrm{65}}}{\tau_{\mathrm{21}}} &= \frac{A_{\mathrm{65}}\nu_{\mathrm{21}}^3\Delta V_{\mathrm{21}}}{A_{\mathrm{21}}\nu_{\mathrm{65}}^3\Delta V_{\mathrm{65}}} \frac{e^{h\nu_{\mathrm{65}}/kT_{\mathrm{65}}}-1}{e^{h\nu_{\mathrm{21}}/kT_{\mathrm{21}}}-1} \frac{N_{\mathrm{6}}}{N_{\mathrm{2}}}.
\end{align}
This expression can be simplified by using the Planck function, assuming that the linewidths are similar, and using that the ratio of the column densities is determined by a Boltzmann distribution:
\begin{align}
\frac{\tau_{65}}{\tau_{21}} &= \frac{A_{65}}{A_{21}} \frac{B_{21}(T)}{B_{65}(T)} \frac{g_6}{g_2}e^{-(E_6-E_2)/kT},
\end{align}
with $g_J = 2J+1$ the degeneracy of level $J$ for CO. The derived optical depths are presented in Fig.~\ref{fig:tau}. The $^{13}$CO $J=2-1$ emission outside the dust cavity is optically thick in both disks as expected, the C$^{18}$O $J=2-1$ transition is moderately optically thick. The optical depths of all transitions in the dust cavity of LkCa15 are upper limits as an upper limit on the integrated C$^{18}$O $J=2-1$ intensity was used. In the case of HD~169142, the optical depth estimates outside 35~AU may be affected by the maximum resolvable scale of these observations.

  \begin{figure*}
        \centering
            \includegraphics[width=\textwidth]{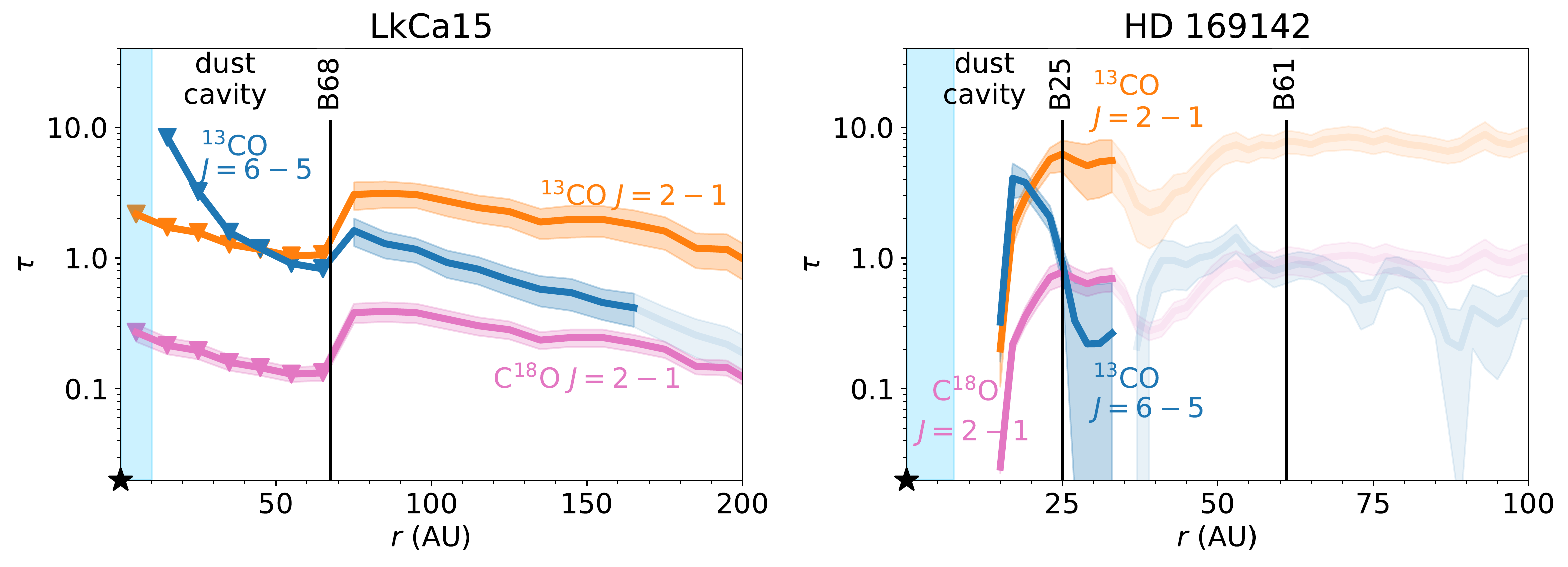}
     
        \caption{Optical depth of the $^{13}$CO $J=6-5$ and $J=2-1$, and the C$^{18}$O $J=2-1$ transitions for LkCa15 (left) and HD~169142 (right). The optical depth of the $J=2-1$ transitions is derived from their respective line ratio and the optical depth of the $^{13}$CO $J=6-5$ transition is derived using the line ratio analysis, see text for details. }
        \label{fig:tau}\end{figure*}

\section{DALI} \label{app:dali} 
\subsection{Model setup}

\begin{figure}
\centering
        \includegraphics[width=\columnwidth]{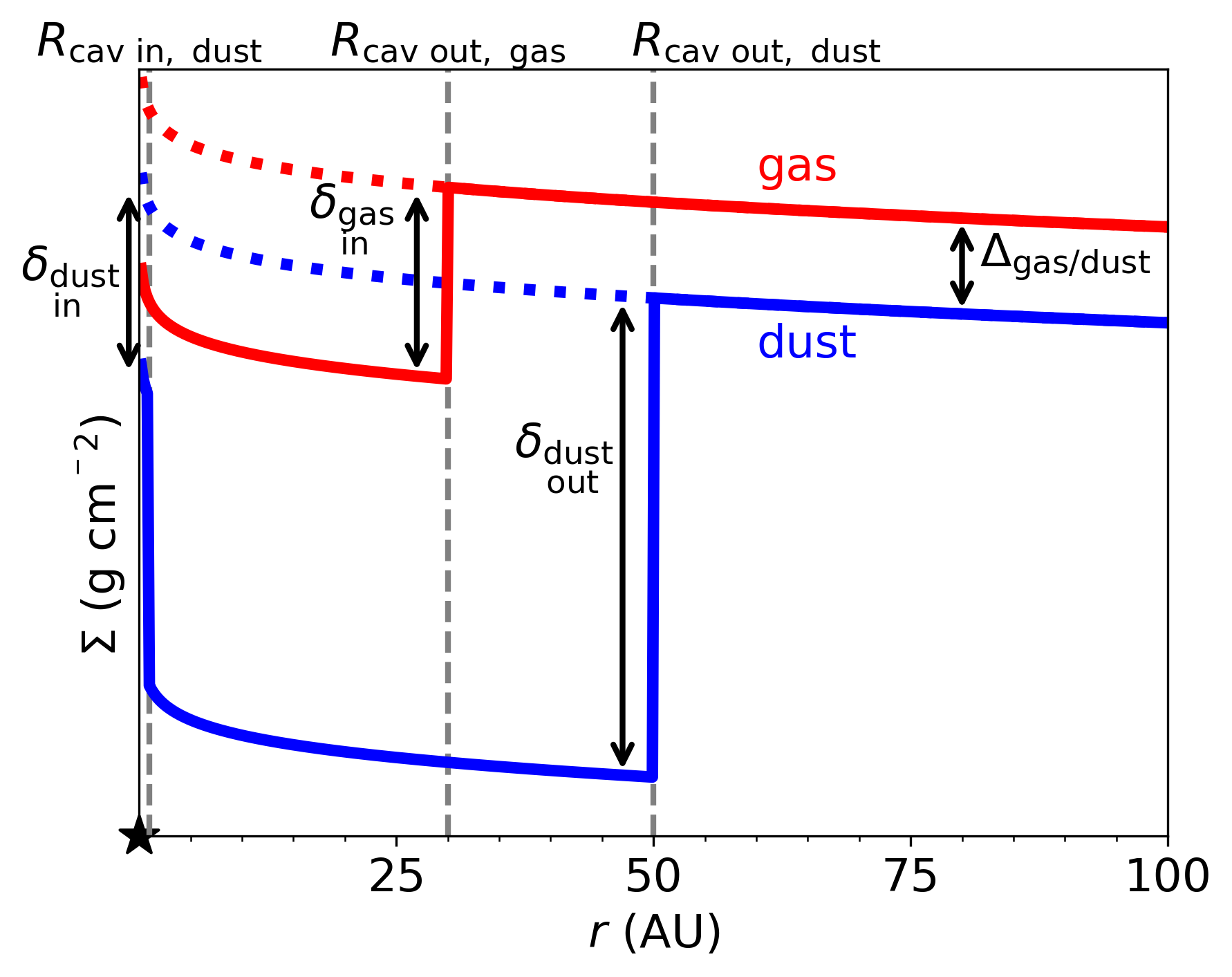}
\caption{Generic structure of the DALI models. The gas surface density is indicated in red and the dust surface density is indicated in blue. }
\label{fig:dali_structure_generic}
\end{figure}

 \begin{figure}
   \centering
  \begin{subfigure}{0.99\columnwidth}
  \centering
  \includegraphics[width=1\linewidth]{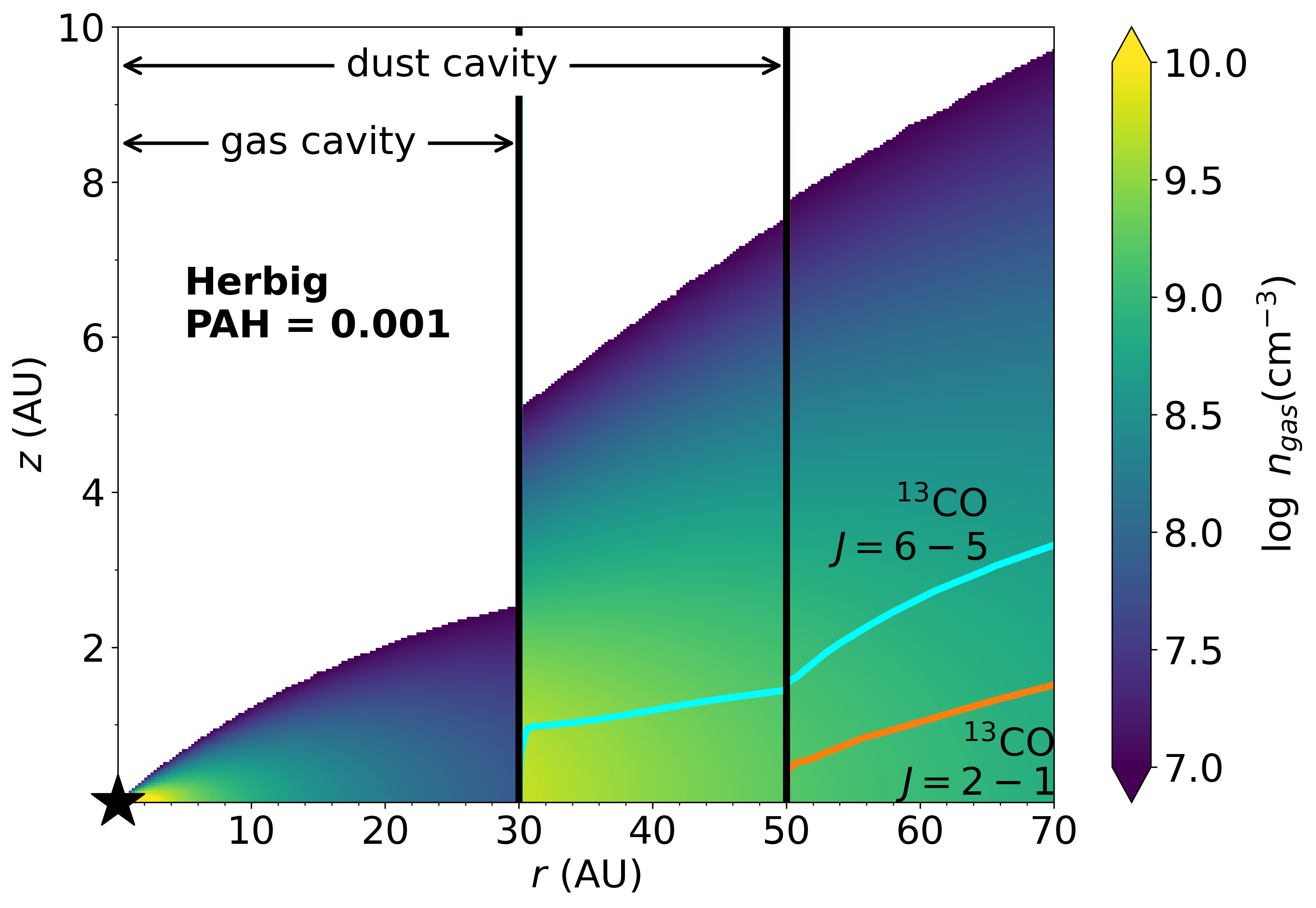}
\end{subfigure}%

\begin{subfigure}{0.99\columnwidth}
  \centering
    \includegraphics[width=1\linewidth]{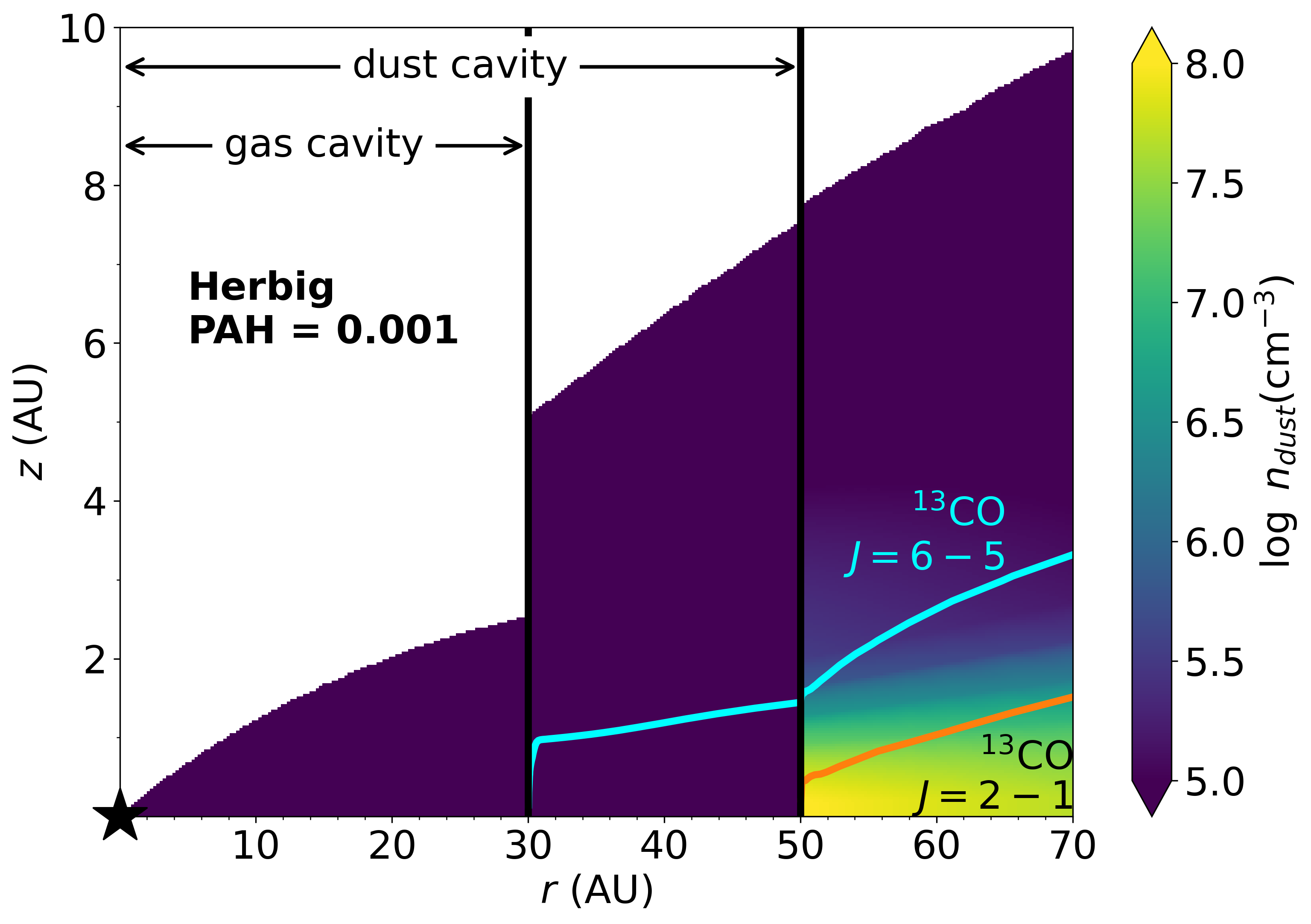}
\end{subfigure}
  \begin{subfigure}{0.99\columnwidth}
  \centering
  \includegraphics[width=1\linewidth]{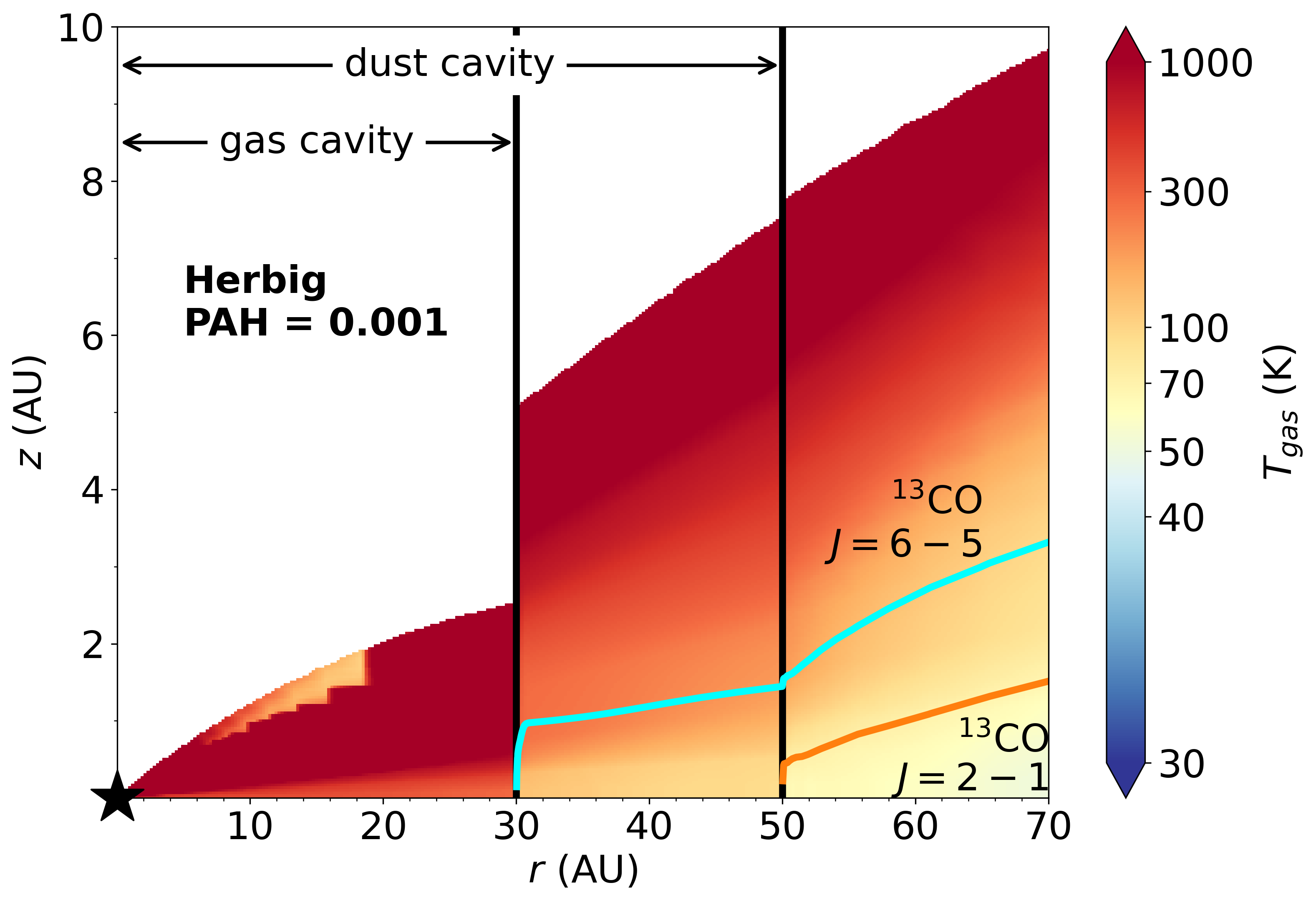}
\end{subfigure}
      \caption{
      The gas number density (top), dust density (middle), and gas temperature (bottom) of our fiducial model for a Herbig star with a factor of 100 drop in gas density inside 30~AU are shown. The cyan and orange contours indicate the $\tau = 1$ surface of the $^{13}$CO $J=6-5$ and $J=2-1$ lines respectively. The $^{13}$CO $J=2-1$ transition is optically thin inside the dust cavity and the $J=6-5$ is optically thin inside the gas cavity, hence these lines trace the midplane at these radii. Note that the dust cavity starts from 50~AU inwards and the gas cavity from 30~AU. Only the regions with a gas density above $10^7$~cm$^{-3}$ are shown. Note that the bottom panel is identical to the middle panel in Fig.~\ref{fig:DALImodelLkCa15T}.}. 
         \label{fig:DALImodelLkCa15n}
   \end{figure}

The exploratory models for the $^{13}$CO $J=6-5$ to $J=2-1$ line ratio discussed in Section~\ref{sec:dalisection} and Appendix~\ref{sec:daliother}, are computed using the thermo-chemical code \texttt{DALI} \citep{Bruderer2009, Bruderer2012, Bruderer2013}. The disk gas and dust structure together with the stellar spectrum are used to calculate the thermal and chemical structure across the disk and predict the flux that can be observed by e.g. ALMA. 

The radial structure of the gas and dust in DALI is based on the self-similar solution to a viscously evolving disk \citep{LyndenBell1974, Hartmann1998}:
\begin{align}
\Sigma_{\mathrm{gas}} (R) = \Sigma_{\mathrm{c}} \left (\frac{R}{R_{\mathrm{c}}} \right )^{-\gamma} \exp \left [-\left (\frac{R}{R_{\mathrm{c}}}\right )^{2-\gamma} \right ],
\end{align}
with $\Sigma_{\mathrm{gas}} (R)$ the gas surface density at radius $R$, $\Sigma_{\mathrm{c}}$, the gas surface density at the characteristic radius $R_{\mathrm{c}}$, and $\gamma$ is the power-law index, similar to the approach of \citet{Andrews2011}. An overview of the gas and dust surface density structure is presented in Fig.~\ref{fig:dali_structure_generic} and an overview of all model parameters is presented in Table~\ref{tab:paramsdali}. The inner radius of the protoplanetary disk is set to the sublimation radius that can be approximated as:
\begin{align}
R_{\mathrm{subl}} &\approx 0.07~\mathrm{AU}\sqrt{\frac{L_{\star}}{\mathrm{L_{\odot}}}},
\end{align}
with $L_{\star}$ the luminosity of the star \citep{Dullemond2001}. 

The vertical structure of the gas in the disk is assumed to follow a Gaussian distribution, where the scale height of the gas at each radius is given determined by the flaring, $\psi$, of the disk:
\begin{align}
h = h_{\mathrm{c}} \left ( \frac{R}{R_{\mathrm{c}}} \right )^{\psi},
\end{align}
with $h$ the scale height angle and $h_{\mathrm{c}}$ the scale height angle at $R_{\mathrm{c}}$. This scale height angle can be converted to a physical height above the disk midplane, $H$, by $H\sim hR$.

The dust is modelled using two populations of dust grains following \citet{Andrews2011}: small grains (5~nm$-$1~$\mu$m) and large grains (5~nm$-$1~mm). Both dust populations follow the MRN distribution ($a^{-3.5}$ with $a$ the grain-size; \citealt{Mathis1977}) and the mass fraction of large grains is controlled by $f_{\mathrm{ls}}$. The small dust grains are assumed to be well-coupled to the gas and hence follow the same radial and vertical distribution. The larger grains are assumed to be settled to the midplane as recent observations have shown (e.g. \citealt{Villenave2020}), therefore the scale height of the large grains is reduced by a factor of $\chi < 1$. Finally, the total dust mass is scaled such that the global gas-to-dust mass ratio in the outer disk, $\Delta_{\mathrm{g/d}}$, matches the ISM value of 100. 

The transition disk structure is parametrized as presented in Fig.~\ref{fig:dali_structure_generic}. The gas cavity is modelled by reducing the gas column density inside $R_{\mathrm{cav\ out, gas}}$ by a factor of $\delta_{\mathrm{gas\ in}} < 1$. The cavity in the dust is modelled in a similar way by a drop in column density controlled by $\delta_{\mathrm{dust\ out}} < 1$ between $R_{\mathrm{cav\ in, dust}}$ to $R_{\mathrm{cav\ out, dust}}$. 
The dusty inner disk between $R_{\mathrm{subl}}$ and $R_{\mathrm{cav\ in, dust}}$ is modelled with a factor of $\delta_{\mathrm{dust\ in}} < 1$ w.r.t. the outer disk. 
The gas cavity is typically smaller than the dust cavity \citep{Bruderer2014, Perez2015, vanderMarel2015, vanderMarel2016}, therefore we use $R_{\mathrm{cav\ out, gas}} \leq R_{\mathrm{cav\ out, dust}}$.

PAHs are assumed to be well mixed with the gas. The abundance of PAHs is modelled as a step-function where PAHs can only exist in the disk regions where the integrated far-UV flux is at most $10^6~G_{\mathrm{0}}$, with $G_{\mathrm{0}}$ the interstellar radiation field. In this region where PAHs are abundant, they provide an additional source of opacity. Thus PAHs act both as a heating agent as well as a source for attenuation of UV radiation. 
A more detailed discussion on the treatment of PAHs w.r.t. the gas can be found in \citet{Bruderer2012, Visser2007}. 

The stellar spectrum is modelled as a 4000~K (10000~K) black body for a T~Tauri (Herbig) star. In the case of the T~Tauri star, the UV-luminosity from accretion onto the star is modelled as an additional 10000~K black body and a mass accretion rate of $10^{-8}~\mathrm{M_{\odot}~yr^{-1}}$, corresponding to an UV-luminosity of 0.15~L$_{\odot}$. This is a typical mass accretion rate for a T~Tauri type star (\citealt{Francis2020} and references therein). 

The temperature structure in DALI is calculated in an iterative manner, starting with the dust temperature. The dust temperature is calculated using Monte Carlo continuum radiative transfer. Initially, the abundances of the species in the chemical network are calculated assuming that the gas temperature is equal to the dust temperature. The chemical network used to calculate these abundances is based on the UMIST 06 network \citep{Woodall2007} and has previously been presented in \citet{Bruderer2013}. No isotope-selective chemistry is included and the network is evaluated assuming steady state. Instead, the isotopologue abundances are scaled to their main isotopologue using their elemental ratios given in Table~\ref{tab:paramsdali}. Finally, the heating and cooling rates, depending on the grain sizes following \citet{Facchini2017, Facchini2018}, are used to calculate the gas temperature independent of the dust temperature. The calculations of the abundances and gas temperatures are repeated until they are consistent. 

Finally, the $^{13}$CO $J=6-5$ and $J=2-1$ transitions are ray-traced using non-LTE excitation adopting the collisional rate coefficients from the LAMDA database \citep{Botschwina1993, Flower1999, Schoier2005}. Our fiducial disk is located at 150~pc and has a face-on orientation.

\begin{table*}
\caption{Other DALI model parameters }             
\label{tab:paramsdali}      
\centering          
\begin{tabular}{p{0.3\columnwidth}p{0.3\columnwidth}p{0.3\columnwidth}p{1.\columnwidth}}     
    \hline\hline
        \centering\arraybackslash Model parameter & \centering\arraybackslash T~Tauri & \centering\arraybackslash Herbig & \centering\arraybackslash Description \\ \hline
        \\[-0.7em]
        \textit{Physical structure} & & & \\
        $R_{\mathrm{subl}}$ & 0.07~AU & 0.2~AU & Sublimation radius, calculated as $0.07$~AU$\sqrt{L_{\star}/\mathrm{L}_{\odot}}$.$^{(1)}$ \\        
        $R_{\mathrm{cav\ in,\ dust}}$ &  \multicolumn{2}{c}{\textbf{1}~AU} & Inner radius of the dust cavity.  \\        
        $R_{\mathrm{cav\ out,\ gas}}$ & \multicolumn{2}{c}{\textbf{30}~AU}& Outer radius of the gas cavity.  \\        
        $R_{\mathrm{cav\ out,\ dust}}$ & \multicolumn{2}{c}{\textbf{50}~AU} & Outer radius of the dust cavity. \\        
        $R_{\mathrm{c}}$ & \multicolumn{2}{c}{\textbf{100}~AU}& Characteristic radius of the surface density profile.  \\        
        $R_{\mathrm{out}}$ & \multicolumn{2}{c}{\textbf{600}~AU}& Outer radius of the disk.  \\
		$\Sigma_{\mathrm{c}}$ & \multicolumn{2}{c}{\textbf{0.28} g$\ \mathrm{cm^{-2}}$}& Sets the gas surface density at the characteristic radius $R_{\mathrm{c}}$. \\    
		$M_{\mathrm{disk}}$ & \multicolumn{2}{c}{\textbf{1.5(-3)}~$\mathrm{M_{\odot}}$}& Mass of the disk. \\   
        $\gamma$ & \multicolumn{2}{c}{\textbf{1}} & Power law index of the surface density profile.  \\
        $h_{\mathrm{c}}$ & \multicolumn{2}{c}{\textbf{0.05}} & Scale height angle at the characteristic radius $R_{\mathrm{c}}$. \\
        $\psi$ & \multicolumn{2}{c}{\textbf{0.05}}& Flaring index of the disk surface density. \\
        PAH abundance & \multicolumn{2}{c}{\textbf{1(-3)}, 1(-1)}& Abundance of PAHs w.r.t. to ISM value, in gas.  \\
        $\delta_{\mathrm{gas,\ in}}$ & \multicolumn{2}{c}{1(-3), \textbf{1(-2)}, 1(-1), 1} & Relative drop in gas density inside the gas cavity ($R_{\mathrm{subl}} < R < R_{\mathrm{cav\ out,\ gas}}$). \\
        $\delta_{\mathrm{gas,\ out}}$ & \multicolumn{2}{c}{\textbf{1}} & Relative drop in gas density inside the dust cavity, but outside the gas cavity ($R_{\mathrm{cav\ out,\ gas}} < R < R_{\mathrm{cav\ out,\ dust}}$).\\
        $\delta_{\mathrm{dust,\ in}}$ & \multicolumn{2}{c}{1(-10), \textbf{1(-4)}}& Relative drop in dust density in the dusty inner disk w.r.t. the outer disk ($R_{\mathrm{subl}} < R < R_{\mathrm{cav\ in,\ dust}}$). \\
        $\delta_{\mathrm{dust,\ out}}$ & \multicolumn{2}{c}{\textbf{1(-10)}}&  Relative drop in dust density inside the dust cavity ($R_{\mathrm{cav\ in,\ dust}} < R < R_{\mathrm{cav\ out,\ dust}}$). \\
        \\[-0.3em]
        \textit{Dust properties} & & &  \\
        $\chi$ & \multicolumn{2}{c}{\textbf{0.2}} & Settling of large grains. \\
        $f_{\mathrm{ls}}$ & \multicolumn{2}{c}{1(-10), 0.85, \textbf{0.98}} & Mass-fraction of grains that is large.  \\
        $\Delta_{\mathrm{gas/dust}}$ & \multicolumn{2}{c}{\textbf{100}} & Global gas-to-dust mass ratio.   \\
        \\[-0.3em]
        \textit{Stellar properties} & & & \\          
        $M_{\star}$ & 0.9~$\mathrm{M_{\odot}}$ & 2.5~$\mathrm{M_{\odot}}$ & Mass of the central star.  \\
        $\dot{M}_{\star}^{(2)}$ & 1(-8)~$\mathrm{M_{\odot}}\ yr^{-1}$  & 0~$\mathrm{M_{\odot}}\ yr^{-1}$& Mass accretion rate of the central star.   \\        
        \\[-0.3em]
        \textit{Stellar spectrum} & & &  \\     
		$L_{\star}$ & 1~$\mathrm{L_{\odot}}$& 10~$\mathrm{L_{\odot}}$ & Luminosity of the central star.   \\
        $L_{\mathrm{X}}$ & \multicolumn{2}{c}{\textbf{1(30)} erg s$^{-1}$} & X-ray luminosity of the central star.  \\
        $T_{\mathrm{eff}}$ & 4000~K & 10000~K & Effective temperature of the central star. \\
        $T_{\mathrm{X}}$ &  \multicolumn{2}{c}{\textbf{7(7)}~K}  & Effective temperature of the X-ray radiation.   \\  
        $\zeta_{\mathrm{c.r.}}$ & \multicolumn{2}{c}{\textbf{5(-17)}~$\mathrm{s^{-1}}$} & Cosmic ray ionization rate. \\ 
        \\[-0.3em]
        \multicolumn{3}{l}{\textit{Observational geometry}}&  \\
        $i$ & \multicolumn{2}{c}{\textbf{0$\degree$ } } & Disk inclination ($0\degree$ is face-on).\\
        $d$ & \multicolumn{2}{c}{\textbf{150}~pc} & Distance to the star. \\ 
        \\[-0.3em]
        \textit{Chemistry} & &&  \\
        H  &  \multicolumn{2}{c}{\textbf{1}}        & Abundance w.r.t. the total number of hydrogen atoms. \\
        He &  \multicolumn{2}{c}{\textbf{7.6(-2)}}  & Abundance w.r.t. the total number of hydrogen atoms. \\
		C  &  \multicolumn{2}{c}{\textbf{1.4(-4)}}  & Abundance w.r.t. the total number of hydrogen atoms. \\
		N  &  \multicolumn{2}{c}{\textbf{2.1(-5)}}  & Abundance w.r.t. the total number of hydrogen atoms. \\
		O  &  \multicolumn{2}{c}{\textbf{2.9(-4)}}  & Abundance w.r.t. the total number of hydrogen atoms. \\
		Mg &  \multicolumn{2}{c}{\textbf{4.2(-7)}}  & Abundance w.r.t. the total number of hydrogen atoms. \\
		Si &  \multicolumn{2}{c}{\textbf{7.9(-6)}}  & Abundance w.r.t. the total number of hydrogen atoms. \\
		S  &  \multicolumn{2}{c}{\textbf{1.9(-6)}}  & Abundance w.r.t. the total number of hydrogen atoms. \\
		Fe &  \multicolumn{2}{c}{\textbf{4.3(-7)}}  & Abundance w.r.t. the total number of hydrogen atoms. \\
        $^{12}$C/$^{13}$C & \multicolumn{2}{c}{\textbf{70}}  & Carbon isotope ratio.  \\
        $^{16}$O/$^{18}$O & \multicolumn{2}{c}{\textbf{560}} & Oxygen isotope ratio. \\ \hline               
\end{tabular}
\tablefoot{$a(b)$ represents $a\times 10^b$. Boldface indicates the fiducial value. $^{(1)}$ Following \citet{Dullemond2001}. $^{(2)}$ The stellar accretion rate is converted to $10^4$~K black body and then added to the stellar spectrum. An accretion rate of $10^{-8}$~$\mathrm{M_{\odot}}\ \mathrm{yr}^{-1}$ corresponds to an UV luminosity of 0.15~L$_{\odot}$. }
    \begin{tablenotes}
      \small
      \item 
    \end{tablenotes}
\end{table*}

\subsection{DALI models for other parameters} \label{sec:daliother}

The effect of the gas cavity depth is discussed in Section~\ref{sec:dalisection}. In this section the effect of other parameters such as the abundance of PAHs, the presence of a dusty inner disk and the mass fraction of large grains is investigated.

The results for the Herbig disk model with a deep ($\times 0.01$) gas cavity and a dusty inner disk\footnote{Other parameters as listed in Table~\ref{tab:paramsdali}.} are presented in Fig.~\ref{fig:dalidiffparams}. The dusty inner disk shields the cavity from UV-radiation, therefore the gas between 30 and 50~AU is warmer in the model without dusty inner disk, which is also reflected in the line ratio. This effect is even more pronounced in the region inside 30~AU, where the gas density is low. 

The importance of PAHs is highlighted in the right two panels of Fig.~\ref{fig:DALImodelLkCa15T} and in the middle panel of Fig.~\ref{fig:dalidiffparams}. Increasing the PAH abundance increases the gas temperature in a thin layer at $z/r\sim0.15$, but it decreases the gas temperature in the midplane and in intermediate layers due to attenuation of UV radiation. Furthermore, the two $^{13}$CO transitions emit from higher disk layers due to this temperature change {color{red} and because CO is abundant to higher disk layers}. Together these two effects result in a lower line ratio of $\sim20$ in the gas cavity, whereas the model with a lower PAH abundance of $10^{-3}$ w.r.t. the ISM predicts a high line ratio inside 30~AU. This is again due to the underlying temperature structure. The high abundance of PAHs in the former model, effectively shields the midplane where the optically thin $^{13}$CO lines are emitting from. The lack of UV radiation is responsible for the low temperatures.  

Finally, the effect of grain growth is investigated by varying the fraction of large grains, $f_{\mathrm{ls}}$, from 98\% in the fiducial model to 85\% and 0\% in the models mimicking disks with smaller grains.
The disk models with many small grains are colder because of the efficient shielding of UV radiation by these grains and PAHs that are abundant to larger scale heights. Therefore, heating through the photoelectric effect on PAH and FUV pumped H$_2$ is less efficient in these regions, causing the line ratio to decrease when more grains are small.

\begin{figure*}
   \centering
  \includegraphics[width=1\linewidth]{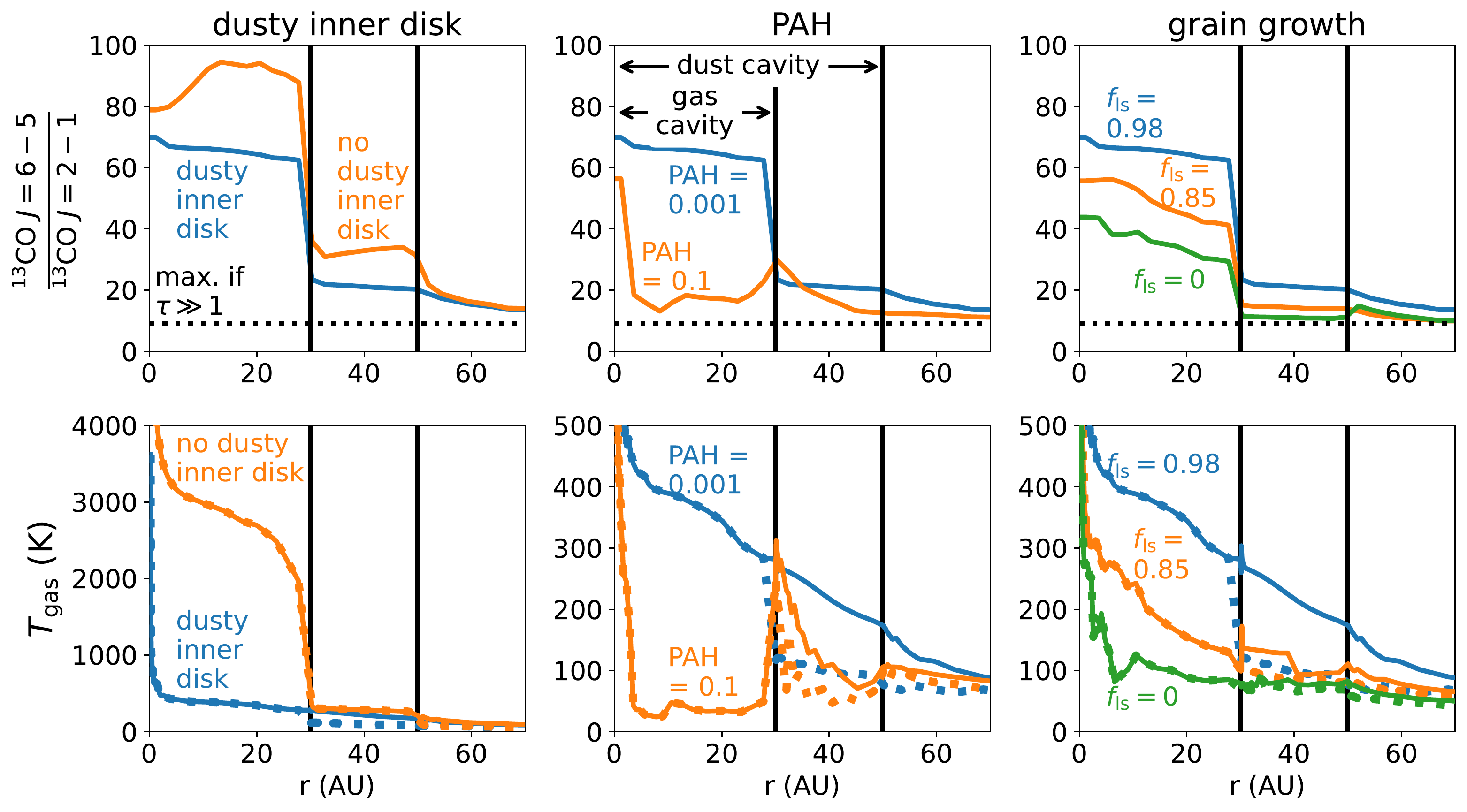}
      \caption{Modelled line ratio for the Herbig disk model (top), where the absence of a dusty inner disk (left), the PAH abundance w.r.t. the ISM (middle) and the fraction of large grains (right) are compared to the fiducial Herbig disk model with a $0.01\times$ drop in gas density inside 30~AU (blue). The bottom row shows the temperature at the $\tau=1$ surface for the $^{13}$CO $J=6-5$ (solid) and $J=2-1$ (dotted) transitions. Both $^{13}$CO lines are (moderately) optically thin inside the gas cavities of these models. Therefore, the midplane temperature is shown in this region. Note the difference in the temperature axes.}
         \label{fig:dalidiffparams}
   \end{figure*}

\end{appendix}

\end{document}